\documentclass[12pt]{article}

\usepackage{amssymb}
\usepackage[dvips]{graphicx}

\def\appendix#1{
  \addtocounter{section}{1}
  \setcounter{equation}{0}
  \renewcommand{\thesection}{\Alph{section}}
  \section*{Appendix \thesection\protect\indent \parbox[t]{11.715cm} {#1}}
  \addcontentsline{toc}{section}{Appendix \thesection\ \ \ #1}
  }

\newcommand {\bd}{\begin{displaymath}}
\newcommand {\ed}{\end{displaymath}}
\newcommand {\eq}{\begin{equation}}
\newcommand {\beq}{\begin{equation}}
\newcommand {\eeq}{\end{equation}}
\newcommand {\beqa}{\begin{eqnarray}}
\newcommand {\eeqa}{\end{eqnarray}}

\newcommand {\tr}{{\rm tr\,}}

\newcommand {\RR}{\mbox{\scriptsize R}}
\newcommand {\II}{\mbox{\scriptsize I}}

\newcommand {\trs}{\mbox{\scriptsize tr}}

\newcommand {\ee}{\mbox{e}}

\newcommand {\dd}{\mbox{d}}

\newcommand {\del}{\partial}

\newcommand {\defeq}{\stackrel{\rm def}{=}}

\newcommand {\vev} [1] {\langle #1 \rangle}
\newcommand{\id}{{1\!\!1}} %% identity operator

\renewcommand{\theequation}{\thesection.\arabic{equation}}

\begin{document}

\setlength{\oddsidemargin}{0cm}
\setlength{\baselineskip}{7mm}

\begin{titlepage}

\baselineskip=14pt

\renewcommand{\thefootnote}{\fnsymbol{footnote}}
\begin{normalsize}
\begin{flushright}
\begin{tabular}{l}
%NBI--HE--02--??\\
DPNU--02--25\\
SUNY--NTG--02/27\\
%hep-th/0108070\\
\hfill{ }\\
August 2002
\end{tabular}
\end{flushright}
  \end{normalsize}

%{}~~\\

%\vskip 1 cm

%\vspace{5mm}

\vspace*{0cm}
    \begin{Large}
       \begin{center}
         {The factorization method for systems with a complex action} \\
         {--- a test in  Random Matrix Theory for finite density QCD ---} \\
%\vspace{1cm}
       \end{center}
    \end{Large}
\vspace{1cm}

\begin{center}
J.\ A{\sc mbj\o rn}$^{1)}$\footnote
           {
e-mail address : ambjorn@nbi.dk},
K.N.\ A{\sc nagnostopoulos}$^{1,2)}$\footnote
            {
e-mail address : konstant@physics.uoc.gr},\\
J.\ N{\sc ishimura}$^{1,3)}$\footnote
           {
e-mail address : nisimura@eken.phys.nagoya-u.ac.jp}
           {\sc and}
           J.J.M.\ V{\sc erbaarschot}$^{4)}$\footnote
           {
e-mail address : verbaarschot@tonic.physics.sunysb.edu}\\
      \vspace{1cm}
        $^{1)}$ {\it The Niels Bohr Institute,}\\
               {\it Blegdamsvej 17, DK-2100 Copenhagen \O, Denmark}\\
      \vspace{3mm}
        $^{2)}$ {\it Department of Physics, University of Crete,}\\
              {\it P.O. Box 2208, GR-71003 Heraklion, Greece}\\
      \vspace{3mm}
        $^{3)}$ {\it Department of Physics, Nagoya University,}\\
              {\it Nagoya 464-8602, Japan}\\
      \vspace{3mm}
        $^{4)}$ {\it Department of Physics and Astronomy, SUNY,}\\
              {\it Stony Brook, NY 11794, USA}\\
\end{center}

\vskip 1 cm

%\hspace{5cm}
\hspace{8cm}

\begin{abstract}
\noindent
Monte Carlo simulations of systems with a complex action 
are known to be extremely difficult.
A new approach to this problem based on a factorization property of
distribution functions of observables has been proposed recently.
%In particular it is free from the so-called overlap problem.
The method can be applied to {\em any} system with a complex action, 
and it eliminates the so-called overlap problem {\em completely}.
We test the new approach in a Random Matrix Theory for finite density
QCD, 
%where exact results are known even for finite matrix size $N$.
where we are able to reproduce the exact results for the 
quark number density.
% at $N=16,24,32,48$.
%These values of $N$ already allow us to extract the
%large-$N$ limit.
The achieved system size is large enough to extract the thermodynamic
limit.
Our results provide a clear understanding 
of how the expected first order phase transition
is induced by the imaginary part of the action.  
\end{abstract}
\vfill
\end{titlepage}
\vfil\eject
\setcounter{footnote}{0}
%keywords for JHEP
%shs     Superstrings and Heterotic Strings 
%mmo     Matrix Models 

\renewcommand{\thefootnote}{\arabic{footnote}}

\baselineskip=18pt

%%%%%%%%%%%%%%%%%%%%%%%%%%%%%

\section{Introduction}

\setcounter{equation}{0}
\renewcommand{\thefootnote}{\arabic{footnote}} 
The (Euclidean) action of many interesting systems in fields ranging from
 condensed matter physics to high-energy physics  has an imaginary part.
%that their action $S$ has an imaginary part $\Gamma$ so that
% \beq
% S = S_0 + i \, \Gamma \ .
% \label{actionS}
% \eeq
%where $S_0$ and $\Gamma$ are real.
Some examples in high-energy physics
are QCD at finite baryon density, Chern-Simons theories,
systems with topological terms (like the $\theta$-term in QCD)
and systems with chiral fermions.
%jac: I substantially changed the sentence below. My believe is that
% if we are not able to calculate things the problem is no simplify
% technical, but somehow we do not have the correct formulation of
% the problem, and, in fact, conceptually do not understand the problem.
While this is not a conceptual problem per se, it severely limits 
the application of Monte Carlo methods, 
which otherwise might provide a powerful tool to understand the properties 
of these systems from first principles.
So far there is no general solution to 
this `complex action problem'.
%, although there exist some methods
%that work for special classes of systems.
%
%(see Refs.\ \cite{signproblem} for recent works).
%
%although there are some special cases in which one can integrate out
%some degrees of freedom to arrive at a system with a real
%action, thereby eliminating the problem completely.

In Ref.\ \cite{sign}
a new Monte Carlo approach to systems with a complex action was proposed.
This method utilizes a simple factorization property of 
distribution functions of observables.
%that occur in the partition function.
Since the property holds quite generally,
the approach can be applied to {\em any} system with a complex action.
Most notably, the method eliminates the so-called {\em overlap
problem}, which occurs when one applies the standard re-weighting 
technique to include the effect of the imaginary part.
% In a concrete example, we will see that the resolution of the
% overlap problem already allows us to achieve the system size
% which is large enough to obtain the thermodynamic limit.
%$\Gamma$.
%This makes Monte Carlo studies of complex action systems
%much easier.
%
% In a concrete example, we will see that the resolution of the
% overlap problem already allows us to achieve the system size
% which is large enough to obtain the thermodynamic limit.
%
Ultimately we hope that this method will enable us, among other things,
to explore the phase diagram of QCD at finite baryon density,
% and/or finite temperature, 
where interesting phases such as a superconducting phase
have been conjectured to appear \cite{bailin,krishna,edward}.    %jac

As a first step toward achieving this goal
we test the new approach in a Random Matrix Theory for finite density
QCD \cite{Stephanov:1996ki}. Random Matrix Theory was 
originally introduced to describe the
spectrum of the Dirac operator at zero chemical potential \cite{SV}  %jac
and has been studied intensively in the
literature. (See Ref.\ \cite{Verbaarschot:2000dy} for a review).
The particular extended model we study can be regarded as a schematic
model for QCD at finite baryon density. 
As one increases the `chemical potential',
the model undergoes a first order phase transition,
where the imaginary part of the action plays a crucial role. 
% $\mu$.
Since it is solvable even for 
finite matrix size $N$ \cite{Halasz:1997he},
it serves as a useful testing ground for
simulation techniques for QCD at finite density.
For instance, the problem with quenched simulations
\cite{Barbour:1986jf}
has been clarified in Ref.\ \cite{Stephanov:1996ki}.
It was also used to test the so-called Glasgow 
method \cite{Barbour:1997bh},
%which was proposed as a technique for simulations
%of QCD at finite baryon density, 
and the source of the problems 
was identified \cite{Halasz:1999gc}.
%In fact it is precisely the overlap problem that makes the Glasgow method
%inefficient.

In this article we apply the new method to both phases of this model
(below and above the critical point)
and obtain the expectation value of the `quark number density'.
The results nicely reproduce the exact results known for finite $N$.
The values of $N$ that are accessible by the new method 
turn out to be large enough to extract the large-$N$ limit. 
%They are much larger than the values obtainable 
%when one applies the standard re-weighting
%technique.
Moreover, our results provide a clear understanding
of how the first order phase transition 
is induced by the imaginary part of the action.
%%%%%%%%%%%%%%%%

The method \cite{Azcoiti:2002vk}
proposed for simulating $\theta$-vacuum like systems
can be regarded as a {\em special} case of the factorization method. A 
simplified version of the method was sufficient because
 the observable was identical to the imaginary part of the action.
%-jac
%and there is no need to calculate the correction factor 
%[$\varphi_i(x)$ in (\ref{factorize})] 
%since it is given trivially.
The essence of the factorization method is that it avoids the
overlap problem by the use of constrained 
simulations.
%-jac  like (\ref{part_pot}) in more general cases.
Results for 2d CP$^3$ and other models are very promising.

The remainder of this article is organized as follows.
In Section \ref{RMTintro}, we define a  Random Matrix Theory for
QCD at finite baryon density 
%the observables,
and review the known exact results.
In Section \ref{problem}, we explain the complex action problem
associated with the standard re-weighting technique. The application
of the factorization
method to a Monte Carlo study of the Random Matrix Model is discussed
in Section \ref{method}.
Section \ref{results} shows the results which nicely reproduce
the exact values for the quark number density.
% The scaling behavior
% of the weight factor in the present model is discussed in section
% \ref{scaling}. We also discuss its  use for extrapolations. 
Section \ref{Summary} is devoted to summary and discussions.

\section{Random Matrix Theory for finite density QCD}
\label{RMTintro}
%\section{Introduction}
\setcounter{equation}{0}
\renewcommand{\thefootnote}{\arabic{footnote}}

The Random Matrix Model we study in this article  is defined by the 
partition function
\beq
Z = \int \dd W \ee ^{- N \, \trs (W^\dag W)} \, \det D  \ ,
\label{rmtdef}
\eeq
where $W$ is a $N \times N$ complex matrix, and 
the $D$ is a $2N \times 2N$ matrix given by
\beq
D = 
\left( 
\begin{array}{cc}
m & i W +  \mu  \\
i W^\dag  + \mu & m
\end{array}
\right) \ .
\label{defD}
\eeq
The parameters $m$ and $\mu$ correspond to the 
`quark mass' and the `chemical potential', respectively. The size of
the matrix can be thought of as the total number of `low-lying' modes 
for given total volume. Since the density of these modes is taken to be
unity, $N$ can be interpreted as the volume of space time.
This model has the global symmetries of the 
Dirac operator of QCD at nonzero baryon chemical potential, where
the matrix $W$ in (\ref{defD}) is a complicated function of
the background gauge field.
Thus the above model can be thought
of as a schematic model for QCD at finite baryon density, where the 
path integral over the gauge field is simply replaced by the
Gaussian integral over $W$.
Interesting observables are
the `chiral condensate' and the `quark number density'
defined by
\beqa
\Sigma &=&  \frac{1}{2N} \, \tr \, (D^{-1}) \\
\nu  & =&  \frac{1}{2N} \, \tr \, ( \gamma_4 D^{-1} ) \ ,
\mbox{~~~~~~~} \gamma_4 = \left( 
\begin{array}{cc}
0 & \id  \\
\id  & 0
\end{array}
\right) \ .
\eeqa

\begin{figure}[htbp]
  \begin{center}
    \includegraphics[height=8cm]{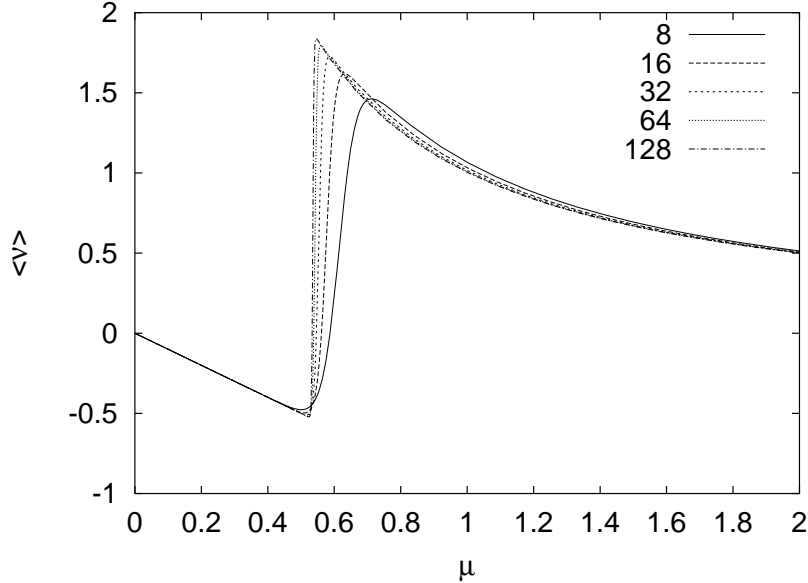}
    \caption{The exact result (\protect\ref{finiteN}) 
for the `quark number density' $\langle \nu \rangle$ is 
plotted as a function of the `chemical potential' $\mu$
for $N=8,16,32,64,128$.
In the $N \rightarrow \infty$ limit, the function develops
a discontinuity at $\mu = \mu_{\rm c}=0.527\cdots$.
}
    \label{fig:nuser}
  \end{center}
\end{figure}

The model was first solved in the large-$N$ limit \cite{Stephanov:1996ki}.
Later it was noticed that the model can be solved even for finite $N$ 
\cite{Halasz:1997he}.
Throughout this paper, we consider the massless case ($m=0$) for simplicity.
Then the partition function can be expressed as
\beq
Z(\mu) = \pi \ee^{\kappa} N ^{-(N+1)} \, N !
\left[ 1+ \frac{(-1)^{N+1}}{N !} \gamma (N+1 ,\kappa)
\right] \ ,
\label{partitionfn}
\eeq
where $\kappa = - N \mu^2$ and $\gamma(n,x)$ is the incomplete 
$\gamma$-function defined by
\beq
\gamma(n,x)=\int_0 ^x \ee^{-t} \, t^{n-1} \, dt \ .
\eeq
From this one obtains the vacuum expectation value 
(VEV) of the quark number density as
\beqa
\langle \nu \rangle
&=& \frac{1}{2N} \frac{\del}{\del \mu} \ln Z(\mu) \\
&=& - \mu \left[ 1 + \frac{\kappa ^N \ee^{-\kappa}}
{ (-1)^{N+1}N\! + \gamma(N+1 , \kappa)   }
\right] \ .
\label{finiteN}
\eeqa
In Fig.\ \ref{fig:nuser} we plot $\langle \nu \rangle$ 
as a function of the chemical potential $\mu$ for $N=8,16,32,64,128$.
The large-$N$ limit of this formula can be taken easily by
applying the saddle-point analysis to the incomplete
$\gamma$-function. We obtain
\beq
\lim_{N\rightarrow \infty}
 \langle \nu \rangle =
\left\{ \begin{array}{ll}
- \mu  & \mbox{for~}\mu < \mu _{\rm c}  \\
1/ \mu  & \mbox{for~}\mu > \mu _{\rm c} \ ,
\end{array} 
\right. 
\label{nq_rmt}
\eeq
where $\mu_{\rm c}$ is the solution to
the equation $1 + \mu^2 + \ln (\mu^2) = 0$,
and its numerical value is given by $\mu_{\rm c} =0.527\cdots$.
We find that the quark number density $\langle \nu \rangle$
has a discontinuity at $\mu = \mu_{\rm c}$.
Thus the schematic model reproduces qualitatively 
the first order phase transition expected to occur
in `real' QCD at nonzero baryon density.

\section{The complex action problem}
\label{problem}
\setcounter{equation}{0}
\renewcommand{\thefootnote}{\arabic{footnote}}

In this section we describe the complex action problem that appears 
in standard Monte Carlo studies of the model (\ref{rmtdef}).
Let us first rewrite (\ref{rmtdef}) as
\beq
Z = \int \dd W \, \ee ^{-S_0 + i \, \Gamma}\ ,
\label{rmtdef2}
\eeq
where we have introduced $S_0$ and $\Gamma$ by
\beqa
S_0 &=& N \, \tr (W ^\dag W) - \ln | \det D | \\
\det D &=&  \ee ^ {i \Gamma}\,  | \det D | \ .
\eeqa
In this form it becomes manifest that the system has a 
{\em complex action}, where the
problematic imaginary part $\Gamma$
is given by the phase of the fermion determinant.
Since the weight $\ee ^{-S_0 + i \, \Gamma}$ in (\ref{rmtdef2})
is not positive definite, 
we cannot regard it as a probability density.
Hence it seems difficult to apply the idea of standard Monte Carlo
simulations,
which reduces the problem of obtaining VEVs to that of taking 
an average over an ensemble generated by the probability density.
One way to proceed is to apply the reweighting method
and rewrite the VEV $\langle \nu \rangle$ as
\beq
\left\langle \nu \right\rangle
= \frac{\left\langle \nu \, \ee ^{i \Gamma }
\right\rangle _{0}}
{\left\langle \ee ^{i \Gamma }
\right\rangle _{0}}  \ ,
\label{VEV}
\eeq
where the symbol $\langle \ \cdot \ \rangle _{0}$ denotes 
a VEV with respect to the so-called {\em phase quenched} 
partition function
\beq
Z_0 = \int \dd W \, \ee ^{- N \, \trs (W^\dag W)} \, | \det D | 
= \int \dd W \, \ee ^{-S_0}\ .
\label{absdef}
\eeq
Since the system (\ref{absdef}) has a positive definite weight,
the VEV $\langle \ \cdot \ \rangle _{0}$ can be evaluated by
standard Monte Carlo simulations.
However, the fluctuations of the phase $\Gamma $ in (\ref{VEV}) 
grows linearly with the size of the matrix $D$,
which is of O($N$).
% in the present case.
Due to huge cancellations,
both the denominator and the numerator of the r.h.s.\ of (\ref{VEV})
vanish as $\ee ^{-{\rm const.} N}$ as $N$ increases,
while the `observables' 
$\ee ^{i \Gamma }$ and $\nu \ee ^{i \Gamma }$
are of O(1) for each configuration.
As a result, 
the number of configurations required to obtain the VEVs with
some fixed accuracy grows as $\ee ^{{\rm const.} N}$ (We remind
the reader that $N$ can be considered as the volume of space time). % jac 
%Note that the exponent is given by the size of the Dirac
%operator, rather than the number of elements in the matrix $W$.
This is the notorious `complex action problem'
(or rather the `sign problem', as we see below).
%which occurs also in many other interesting systems 
%(See Refs.\ \cite{signproblem} for recent works).
See Refs.\ \cite{Toussaint:1989fn} for simulation results %jac
for `real'  finite density QCD obtained by this reweighting technique.

In fact we may simplify the expression (\ref{VEV}) slightly
by using a symmetry.
We note that the fermion determinant $\det D$,
as well as the observable $\nu$,
becomes complex conjugate under the transformation
\beq
W \mapsto -  W \ , 
\label{transf}
\eeq
while the Gaussian action remains invariant.
From this we find that
\beqa
\label{sum_nu}
&~& \langle \nu \rangle = \langle \nu_{\rm R} \rangle
+ i \, \langle \nu_{\rm I} \rangle  \\
&~&  \langle \nu_{\rm R} \rangle =
\frac{\left\langle \nu_{\rm R} \cos \Gamma 
\right\rangle _{0}}
{\left\langle \cos \Gamma 
\right\rangle _{0}}  ~~~;~~~
\langle \nu_{\rm I} \rangle =
i \, \frac{\left\langle \nu_{\rm I} \sin \Gamma 
\right\rangle _{0}}
{\left\langle \cos \Gamma 
\right\rangle _{0}}  \ ,
\label{reweight}
\eeqa
where $\nu_{\rm R}$ and $\nu_{\rm I}$ denote the real part
and the imaginary part of $\nu$, respectively.
%As we see below, the VEV $\langle \nu_{\rm I} \rangle$ is actually
%pure imaginary.
In Eq.\ (\ref{reweight}) the problem takes the form of the `sign
problem', since $\, \cos  \Gamma \, $ and $\, \sin  \Gamma \, $ 
flip their sign violently
as a function of the configuration $W$.
Note that both terms in the r.h.s.\ of (\ref{sum_nu}) are real,
meaning in particular that their sum $\langle \nu \rangle$ is also
real.

The model (\ref{absdef}) is solvable in the 
large-$N$ limit \cite{Stephanov:1996ki}.
For $m=0$ one obtains
\beq
\lim_{N\rightarrow \infty}
\langle \nu \rangle_0 = 
\left\{ \begin{array}{ll}
\mu  & \mbox{for~}\mu < 1  \\
1/ \mu   & \mbox{for~}\mu > 1 \ .
\end{array} 
\right. 
\label{nq_abs}
\eeq
In this case the VEV of the quark number density is a continuous function
of the chemical potential $\mu$ unlike in (\ref{nq_rmt}).
This remarkable difference between (\ref{nq_rmt}) and (\ref{nq_abs})
is precisely due to the imaginary part $\Gamma$ of the action.
%If we do not include its effects properly, 
%we cannot observe the first order phase transition
The two results agree trivially at $\mu = 0$, but interestingly
they also agree at $\mu > 1$.
This is because the eigenvalues of $W$ are located inside the
complex unit circle \cite{Halasz:1999gc}. If $\mu > 1$,
the effect of the fermion determinant is $1/N$
suppressed, and the quenched approximation,
as well as the phase quenched approximation,
becomes exact in the thermodynamic limit.
% is not close to %jac
%$- i \lambda$ or $- i \lambda^*$, where
%$\lambda$ is an eigenvalue of $W$, 
%at large $\mu$, perturbative expansions with respect
%to $1/\mu$ becomes valid, and then according to the usual argument,
%the effects of the fermions are suppressed by $1/N$ at large $N$.
Note also that the symmetry under (\ref{transf}) implies
\beq
\langle \nu_{\rm I}\rangle _0 = 0 ~~~;~~~
\langle \nu _{\rm R} \rangle _0 =\langle \nu \rangle _0 \ .
\label{RI0}
\eeq
%due to the symmetry under (\ref{transf}).

We can also predict $\langle \nu_{\rm R} \rangle$ and
$\langle \nu_{\rm I} \rangle$ separately for the unquenched model
in the large $N$ limit.
For that we note that $\nu(-\mu) = - \nu(\mu)^*$. Therefore we have
\beqa
\langle \nu_{\rm R} \rangle
= \frac 12 \Bigl\{ \langle\nu(\mu)\rangle
- \langle \nu(-\mu) \rangle \Bigr\},\qquad
\langle \nu_{\rm I} \rangle
= \frac 1{2i }\Bigl\{ \langle \nu(\mu)\rangle 
+\langle \nu(-\mu) \rangle \Bigr\} \ ,
%,\qquad
\eeqa
where in the calculation of $\langle \nu(-\mu) \rangle$ 
the sign of the chemical potential 
in the fermion determinant is not changed.
In calculating $\langle \nu(\mu) \rangle$,
the fermion determinant $\det D(\mu)$ cancels the poles of $\nu(\mu)$.
This is responsible for the difference between
quenched and unquenched results for $\mu < 1$.
However, the cancellation of the poles does not occur
in calculating $\langle \nu(-\mu) \rangle$.
Therefore we expect that
$\langle \nu(-\mu) \rangle$ coincides with the corresponding
quenched result $\langle \nu(-\mu) \rangle_0$ even for $\mu < 1$.
This implies in particular that for $\mu < \mu_{\rm c}$ we obtain
the results $\langle \nu_{\rm R} \rangle = 0$ 
and $i \, \langle \nu_{\rm I} \rangle =
-\mu$ in sharp contrast to the quenched result (\ref{RI0}).
For $\mu > 1$, on the other hand,
both $\langle \nu_{\rm R} \rangle$ and $\langle \nu_{\rm I} \rangle$
agree with the corresponding quenched results.
These results are indeed observed in our Monte Carlo simulations,
which shall be discussed in Section \ref{results}.

\section{The factorization method}
\label{method}
\setcounter{equation}{0}
\renewcommand{\thefootnote}{\arabic{footnote}}

\subsection{The basic formulae}

In this section, we explain how the factorization method \cite{sign}
can be used to obtain the VEVs $\langle \nu_{\rm R} \rangle$
and $\langle \nu_{\rm I} \rangle$.
The fundamental objects of the method
are the distribution functions
\beq
\rho_{\rm R}(x) \defeq \langle \delta (x - {\rm Re}(\nu)  )\rangle 
\quad {\rm and}\quad
\rho_{\rm I}(y) \defeq \langle \delta (y - {\rm Im}(\nu)  )\rangle 
\label{def_rho}
\eeq
associated with the complex valued $\nu$ .
In a unified notation that will be used below these equations can
be rewritten as
\beq
\rho_i(x) \defeq \langle \delta (x - \nu_i  )\rangle 
~~~~~~i={\rm R,I} \ .
\eeq
In terms of these functions,
the VEVs of $\nu_i$ can be expressed as
\beq
\langle \nu_i \rangle = \int _{-\infty} ^{\infty}
 \dd x \, x \, \rho_i (x)
%~~~~~~i={\rm R,I}
 \ .
\label{formula}
\eeq
%where we have used the symmetry (\ref{rhoI_asym})
%in the second line of Eq.\ (\ref{formulaI}).

What is essential for the method are the constrained partition functions
\beq
Z_i(x) = \int \dd W \, \ee^{-S_0} \, \delta(x-\nu_i) 
\label{cnstr_part}
\eeq
and the average of the phase $\ee ^{i \, \Gamma}$ 
with respect to these partition functions
\beq
\varphi_i(x) \defeq \langle \ee ^{i\, \Gamma} \rangle_{i,\, x} \ .
\eeq
%where $\langle \ \cdot \ \rangle_{i,\, x}$ denotes
%averaging with respect to the partition functions $Z_i(x)$. 
If we also introduce the distribution functions for
the phase quenched model by
%with respect to the phase quenched partition function by
\beq
\rho^{(0)}_i (x) \defeq \langle \delta (x - \nu_i )\rangle_0
%~~~~~~i={\rm R,I}
 \ ,
\eeq
the distribution functions for the unquenched model trivially factorize as
\beq
\rho_i(x) = \frac{1}{C} \, \rho^{(0)}_i (x) \, \varphi _i(x)  
~~~~~~i={\rm R,\,I} \ ,
\label{factorize}
\eeq
where the normalization constant  $C$ is  given by
\beq
C \defeq \langle \ee ^{i \, \Gamma}\rangle _0  \ .
\label{Cdef}
\eeq
Plugging (\ref{factorize}) into (\ref{formula}), we get
\beq
\langle \nu_i \rangle =
\frac{1}{C} 
 \int _{-\infty} ^{\infty}
 \dd x \, x \, \rho_i^{(0)} (x) \, \varphi_i(x) \ .
\eeq
Note also that Eq.\ (\ref{factorize}) implies that
the constant $C$ can be written as
\beq
C = \int _{-\infty} ^{\infty} \dd x 
\, \rho^{(0)} _{\rm R} (x) \,  \varphi _{\rm R} (x) \ .
\label{C_new0}
\eeq
Thus the VEVs $\langle \nu_i \rangle$ 
($i={\rm R,I}$)
can be expressed solely in terms of the functions
$\rho_i^{(0)} (x)$ and $\varphi_i(x)$ ($i={\rm R,\,I}$),
which can be calculated by standard Monte Carlo simulations.
Clearly the derivation presented here is quite general,
and we can obtain similar formulae in any complex action system.

In the present case, 
we can slightly simplify the formulae due to the symmetry under
(\ref{transf}). Using the properties
\beqa
\varphi_{\rm R} (x)^* &=& \varphi_{\rm R}  (x) \ ,\\
\varphi_{\rm I} (x) ^* &=& \varphi_{\rm I} (-x) \ ,\\
\rho^{(0)} _{\rm I} (-x) &=& \rho^{(0)} _{\rm I} (x)  \ ,
\eeqa
which follow from the symmetry, we arrive at
\beqa
\label{simpleformula0}
\langle \nu_{\rm R} \rangle &=&
\frac{1}{C} 
 \int _{-\infty} ^{\infty}
 \dd x \, x \, \rho_{\rm R}^{(0)} (x) \, w_{\rm R}(x) \ ,\\
\langle \nu_{\rm I} \rangle &=&
%i \,
  \frac{2 \, i}{C} 
\int  _{0} ^{\infty}
\dd x \, x \, \rho ^{(0)}_{\rm I} (x) \, w_{\rm I} (x)
\label{simpleformula} \ ,\\
C &=& \int _{-\infty} ^{\infty} \dd x 
\, \rho^{(0)} _{\rm R} (x) \,  w _{\rm R} (x) \ ,
\label{C_new}
\eeqa
where the weight functions $w_i(x)$ are defined by
%we have introduced
\beqa
\label{wRcos}
w_{\rm R}  (x) &\defeq& \langle  \cos \Gamma \rangle_{\RR , x }  \ ,\\
w_{\rm I}(x) &\defeq& \langle \sin \Gamma\rangle_{\II , x}  \ .
\label{wIsin}
\eeqa
Thus the problem reduces to the calculation of the four real
functions $\rho^{(0)} _i (x)$, $w_i(x)$ ($i={\rm R},{\rm I}$),
which we will now discuss.

\subsection{Monte Carlo evaluation of $\rho^{(0)} _i (x)$ and $w_i(x)$}
\label{MCsim}

In order to obtain $w_i(x)$ ($i={\rm R,\,I}$),
we need to simulate (\ref{cnstr_part}).
In practice, we simulate a partition function where
the $\delta$-function is replaced by a sharply peaked potential 
\beq
Z_{i,V}
 =  \int \dd W
 ~ \ee^{-S_0 }~ \ee ^{- V(\nu_i) }  \ .
\label{part_pot}
\eeq
%where $V$ is some potential introduced
%to constrain $\nu_i$ ($i = {\rm R, I}$) to some value
%allowing small fluctuations.
In this study we use a Gaussian potential
\beq
V(x) = \frac{1}{2} \gamma (x - \xi )^2   \ ,
\label{pot}
\eeq
where $\gamma$ and $\xi$ are real parameters.
The results are insensitive to the choice of $\gamma$
as far as they are big enough (we used $\gamma = 1000.0$).
Let us denote the VEV associated with 
the partition function (\ref{part_pot})
as $\langle {\cal O}\rangle _{i,V}$.
The expectation value $\langle \cos \Gamma \rangle _{{\rm R},V}$
represents the value of $w_{\rm R} (x)$ at 
$x = \langle \nu_{\rm R} \rangle _{{\rm R},V}$,
while
the expectation value $\langle \sin \Gamma \rangle _{{\rm I},V}$
represents the value of $w_{\rm I} (x)$ at 
$x = \langle \nu_{\rm I} \rangle _{{\rm I},V}$.

In fact we can also obtain the functions
$\rho_i^{(0)} (x)$ from the same simulation (\ref{part_pot}).
For that we note that the distribution of $\nu _ i$
in the system (\ref{part_pot}) is given by
\beq
\rho _{i,V} (x) \defeq
\langle \delta (x - \nu_i  )\rangle _{i,V} \propto
\rho ^{(0)}_i (x) \, \ee^{ - V ( x )} \ ,
\label{part_pot_rewritten}
\eeq
which typically has a peak.
The position of the peak, which we denote as $x=\tilde{x}$,
is given by the solution to
\beq
0 = \frac{d}{d x} \ln \rho _i (x)  = 
f ^{(0)} _i (x) -  V'(x )   \ ,
\eeq
where we have introduced
\beq
f_i^{(0)} (x)  \defeq  \frac{d}{d x} \ln \rho _i ^{(0)} (x) \ .
\label{def_f}
\eeq
Therefore, 
the quantity $V' (\tilde{x})$ 
represents the value of $f  ^{(0)} _i (x)$
at $x=\tilde{x}$.
Since the value of  $\gamma$ is taken to be large, the peak is sharp and 
we can safely approximate $\tilde{x}$ by 
the VEV $\langle \nu_i \rangle_{i,V}$.
Once we obtain $f  ^{(0)} _i (x)$ for various $x$,
we can calculate $\rho _i ^{(0)} (x)$ by integrating (\ref{def_f})
and exponentiating,
where the integration constant can be determined by the normalization
of the distribution $\rho _i ^{(0)} (x)$. 
If the distribution $\rho^{(0)}_i(x)$ would be Gaussian
(as it is the case near its peak; see Appendix),
the above procedure is exact even for finite $\gamma$.

Monte Carlo simulation of (\ref{part_pot}) can be
performed by using the Hybrid Monte Carlo algorithm
in much the same way as in Refs.\ \cite{AABHN}.
%(The extra factor $\ee ^{- V( \lambda_i) }$ in (\ref{part_pot}) 
%should be incorporated in the Molecular Dynamics evolution,
%as well as in the Metropolis procedure.)
The required computational effort for the present model is O($N^3$).

\subsection{The virtues of the method}
\label{virtue}

Comparing (\ref{simpleformula0}), (\ref{simpleformula}) and
(\ref{C_new}) with Eq.\ (\ref{reweight}), we notice that
\beqa
\label{rel0}
\langle \nu_{\rm R} \cos \Gamma \rangle_0 &=&
 \int _{-\infty} ^{\infty}
 \dd x \, x \, \rho_{\rm R}^{(0)} (x) \, w_{\rm R}(x) \ ,\\
\langle \nu_{\rm I} \sin \Gamma \rangle _0 &=&
%i \,
2 \int  _{0} ^{\infty}
\dd x \, x \, \rho ^{(0)}_{\rm I} (x) \, w_{\rm I} (x) \ , \\
\label{rel}  
% \\
% \langle  \cos \Gamma \rangle_0 &=&
%  \int _{-\infty} ^{\infty}
%  \dd x \,  \rho_{\rm R}^{(0)} (x) \, w_{\rm R}(x) 
% \label{rel2}
\langle  \cos \Gamma \rangle_0 &=&
  \int _{-\infty} ^{\infty}
  \dd x \,  \rho_{\rm R}^{(0)} (x) \, w_{\rm R}(x)  \ .
\label{rel2}
\eeqa
Thus the new method as it stands simply amounts to
using the standard reweighting formula (\ref{reweight}),
but calculating each VEV 
by using (\ref{rel0}), (\ref{rel}) and (\ref{rel2}).
We now  explain the virtues of the present method.
% advantage of calculating the functions
% $\rho_i^{(0)} (x)$ and $w_i(x)$
% instead of calculating the VEVs directly.

If we are to obtain the VEVs 
%l.h.s.\ of these formulae
by directly simulating the system (\ref{absdef}),
for most of the time we sample configurations
whose $\nu_i$ takes a value close to the peak
of $\rho_i^{(0)} (x)$.
However, from the r.h.s.\ of (\ref{rel0})$\sim$(\ref{rel2})
it is clear that 
we have to sample configurations whose $\nu_i$ takes a value 
where $\rho_i^{(0)} (x) |w_i(x)|$ 
becomes large, in order to obtain the VEVs accurately.
In general these two regions of configuration space
have little overlap, which makes the sampling ineffective.
This is the overlap problem.
Since the overlap becomes exponentially small as the system size
increases, this composes some portion of the complex action problem.
The use of (\ref{rel0})$\sim$(\ref{rel2})
and calculating the relevant functions as explained in the previous
section avoids this problem
by `forcing' the simulation to sample the important region.
%as required.
%one wishes.

%%%%%%%%%%%%%%%%

The knowledge of the weight factor $w_i(x)$,
which is provided by the present approach, allows
us to probe directly the effect of 
the imaginary part $\Gamma$ on the observable of our concern.
We can understand which values of the observable are enhanced
or suppressed by the effect of $\Gamma$.
On the other hand, the standard reweighting technique simply gives
the integrals on the l.h.s.\ of (\ref{rel0})$\sim$(\ref{rel2}),
from which we can hardly imagine how they resulted 
from the effect of $\Gamma$.
In the particular model we are studying, the imaginary part $\Gamma$
has a dramatic effect on the VEV of the quark number
as we discussed in Section \ref{problem}.
We will see in the next section that the weight factor $w_i(x)$
indeed provides a clear understanding of this phenomenon.

% for the distribution function of the baryon number.
As we also see in the next section,
a Monte Carlo calculation of the weight factor $w_i(x)$
becomes increasingly difficult with the system size.
In that sense the 
%the method does not solve the
complex action problem is still there.
%completely.
%although it eliminates the overlap problem which is 
This should be contrasted with the meron-cluster algorithm \cite{meron},
which has been applied to a special class of complex action systems
with computational efforts increasing at most by some power
of the system size.
However, as mentioned, the factorization method 
eliminates the overlap problem and this is
a substantial step forward which allows us to 
get closer to the thermodynamic limit. Indeed,
we are able to obtain the thermodynamic limit of the `quark number
density' in this random matrix model with modest 
computer resources.
See also Ref.\ \cite{Fodor:2001au} for an idea to ameliorate 
the overlap %-jac
problem in  `real' QCD at finite baryon density
by interpolations in the ($\mu$, $T$) plane 
(notice however \cite{critique}).                           %jac

%It might often be possible to extend the application of
%the factorization method.
Using the generic scaling properties 
of the weight factor $w_i(x)$,
%representing the effect of the imaginary part
%Using these properties, 
one may extrapolate
the results obtained by direct Monte Carlo evaluations
to larger system size.
Such an extrapolation is expected to be particularly useful
in cases where the distribution function turns out to be positive 
definite.
In those cases we can actually even {\em avoid} using the reweighting 
formula (\ref{reweight}) by
reducing the question of obtaining the expectation value
to that of finding the minimum of the free energy,
which is (minus) the {\it log} of the distribution function.
Here, the error in obtaining the scaling function
propagates to the final result without significant magnifications.
Therefore, the extrapolation can be a powerful tool to 
probe the thermodynamic limit from the accessible system size.
Indeed such a technique has been used in Ref.\ \cite{sign}
to discuss
the dynamical generation of four-dimensional space-time
in a nonperturbative formulation \cite{IKKT}
of type IIB superstring theory in ten dimensions.
%the IIB matrix model \cite{IKKT}, 
%a conjectured nonperturbative definition of type IIB superstring theory.

% The hope is that
% due to the absence the overlap problem,
% the thermodynamic properties can be already study
% at the system size accessible with the available computer resources.
%
% One can apply it essentially to any complex action system.
% If the gain that arises from eliminating the overlap problem
% is large enough, one can study the thermodynamic properties
% at the system size accessible with the available computer resources.

%%%%%%%%%%%%%%%%%%%%%%%

%\newpage 

%\vspace*{1cm}

\section{Reproducing exact results by the new method}
\label{results}
\setcounter{equation}{0}

In this section we
apply the factorization method to 
the model (\ref{rmtdef})
as described in the previous section
and show that  it reproduces the known 
results $\langle \nu \rangle$ given by (\ref{finiteN})
for arbitrary $N$.
Let us recall the dramatic difference between
(\ref{nq_rmt}) and (\ref{nq_abs}).
% at $\mu < \mu_{\rm c}$.
It is particularly interesting to see how this occurs
due to the effects of $\Gamma$ in the present approach.
First we focus on two values of $\mu$, 
$\mu = 0.2$ and $\mu = 1.0$, which are on opposite  
sides of the first order 
phase transition point $\mu = \mu_{\rm c}=0.527 \cdots$.
% We are able to obtain results up to large $N$, which allows
% us to extract the large $N$ limit.
% Then we turn to the critical regime 

\begin{table}[ht]
\begin{center}
\begin{tabular}{|c | c | c | c | c | c | }
\hline\hline
$\mu$&$N$ &$\vev{\nu_{\rm R}}$&$i \, \vev{\nu_{\rm I}}
$&$\vev{\nu}$&$\vev{\nu}$~(exact)\\
\hline
 0.2 & 8  & 0.0056(6) & -0.1970(5)  & -0.1915(7) & -0.20000\ldots \\
 0.2 & 16 & 0.0060(4) & -0.1905(13) & -0.1845(13)& -0.20000\ldots \\
 0.2 & 24 & 0.0076(9) & -0.1972(14) & -0.1896(17)& -0.20000\ldots  \\
 0.2 & 32 & 0.0021(8) & -0.1947(19) & -0.1927(25)& -0.20000\ldots  \\
 0.2 & 48 & 0.0086(37)& -0.2086(54) & -0.2000(88)& -0.20000\ldots  \\
\hline
 1.0 &  8 & 0.8617(10)&  0.1981(13) &  1.0598(12)&1.066501$\ldots$\\
 1.0 & 16 & 0.8936(2) &  0.1353(6)  &  1.0289(5) &1.032240$\ldots$\\
 1.0 & 32 & 0.9207(1) &  0.0945(2)  &  1.0152(3) &1.015871$\ldots$ \\
\hline\hline
\end{tabular}
\end{center}
\caption{Results of the analysis of $\langle \nu \rangle $
described in the text.
Statistical errors computed by the jackknife method are shown.
The last column represents the exact result 
(\protect\ref{finiteN}) for $\vev{\nu}$ at each $\mu$ and $N$.
For $\mu=0.2$ eq.\ (\protect\ref{finiteN})
yields
$\vev{\nu}=-0.2$ with an accuracy better than $1$ part in
$10^{-9}$. 
}
\label{t:1}
\end{table}

\subsection{$\mu < \mu_{\rm c}$}

\begin{figure}[htbp]
  \begin{center}
    \includegraphics[height=8cm]{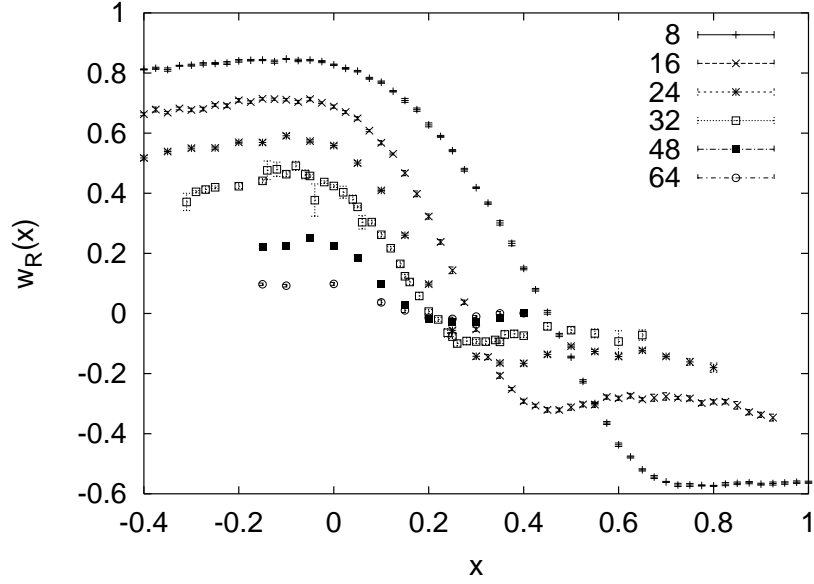}
    \caption{The weight factor $w_{\rm R}(x)$ is plotted against $x$
for $N=8,16,24,32,48,64$ at $\mu = 0.2$.}
    \label{fig:wR}
  \end{center}
\end{figure}

\begin{figure}[htbp]
  \begin{center}
    \includegraphics[height=8cm]{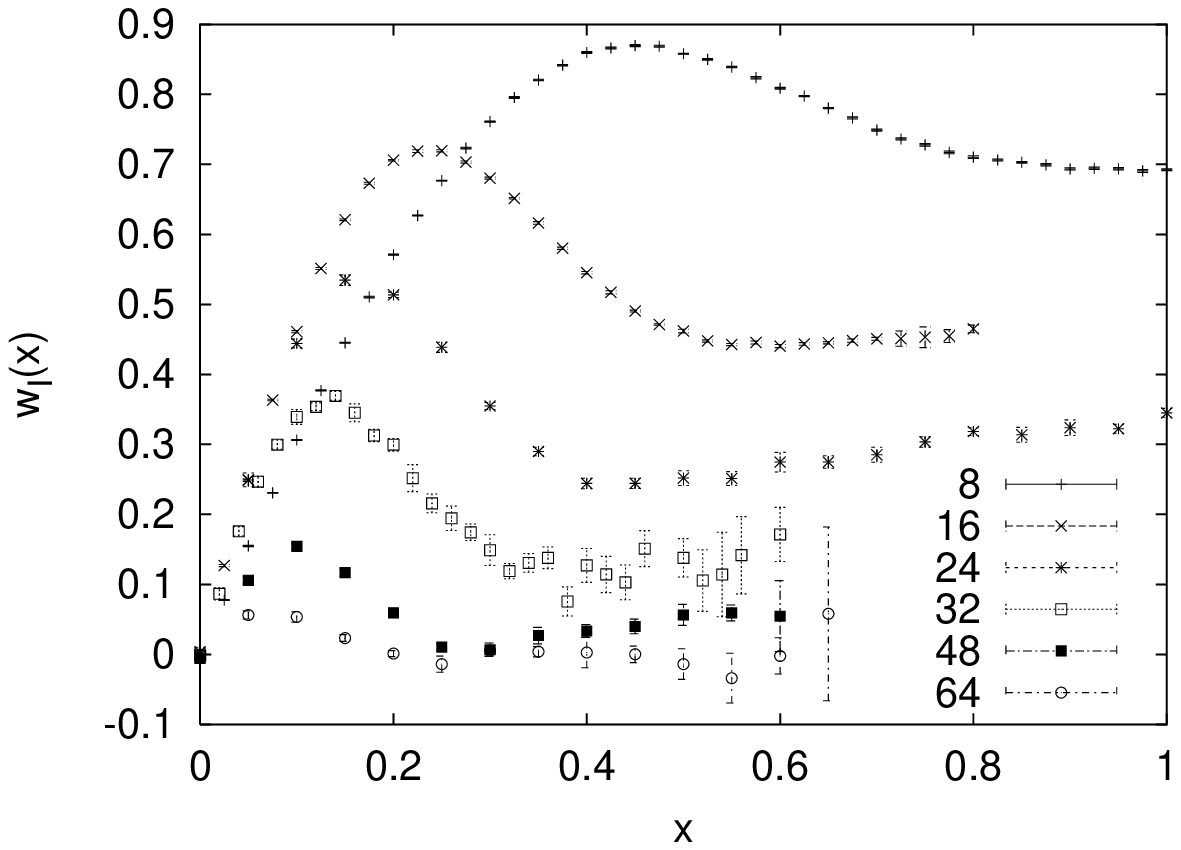}
    \caption{The weight factor $w_{\rm I}(x)$ 
is plotted against $x$
for $N=8,16,24,32,48,64$ at $\mu = 0.2$.}
    \label{fig:wI}
  \end{center}
\end{figure}

\begin{figure}[htbp]
  \begin{center}
    \includegraphics[height=8cm]{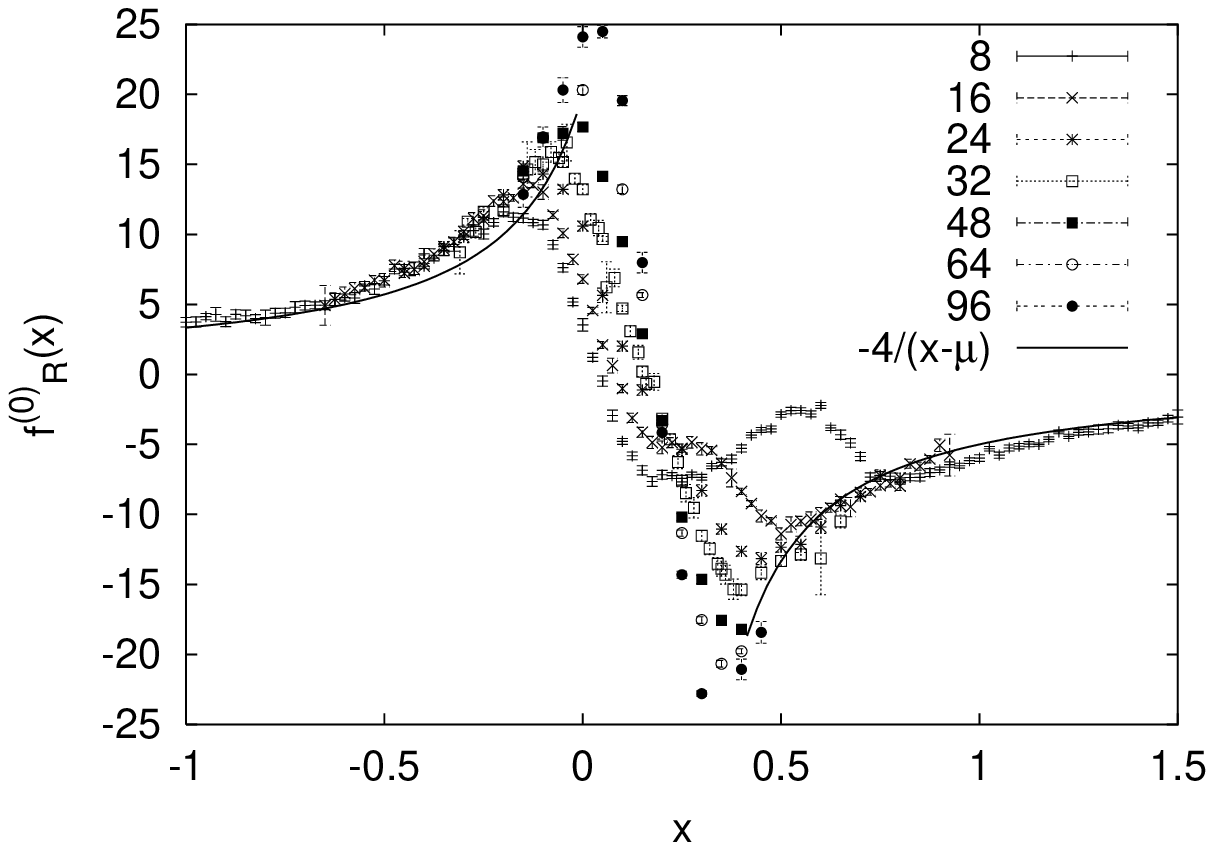}
    \caption{The function $f _{\rm R} ^{(0)} (x)$ is plotted
%against $x$
for $N=8,16,24,32,48,64,96$ at $\mu = 0.2$.
The solid line represents the asymptotic behavior (\ref{f0R_fit})
discussed in the Appendix.
}
    \label{fig:f0R}
  \end{center}
\end{figure}

\begin{figure}[htbp]
  \begin{center}
    \includegraphics[height=8cm]{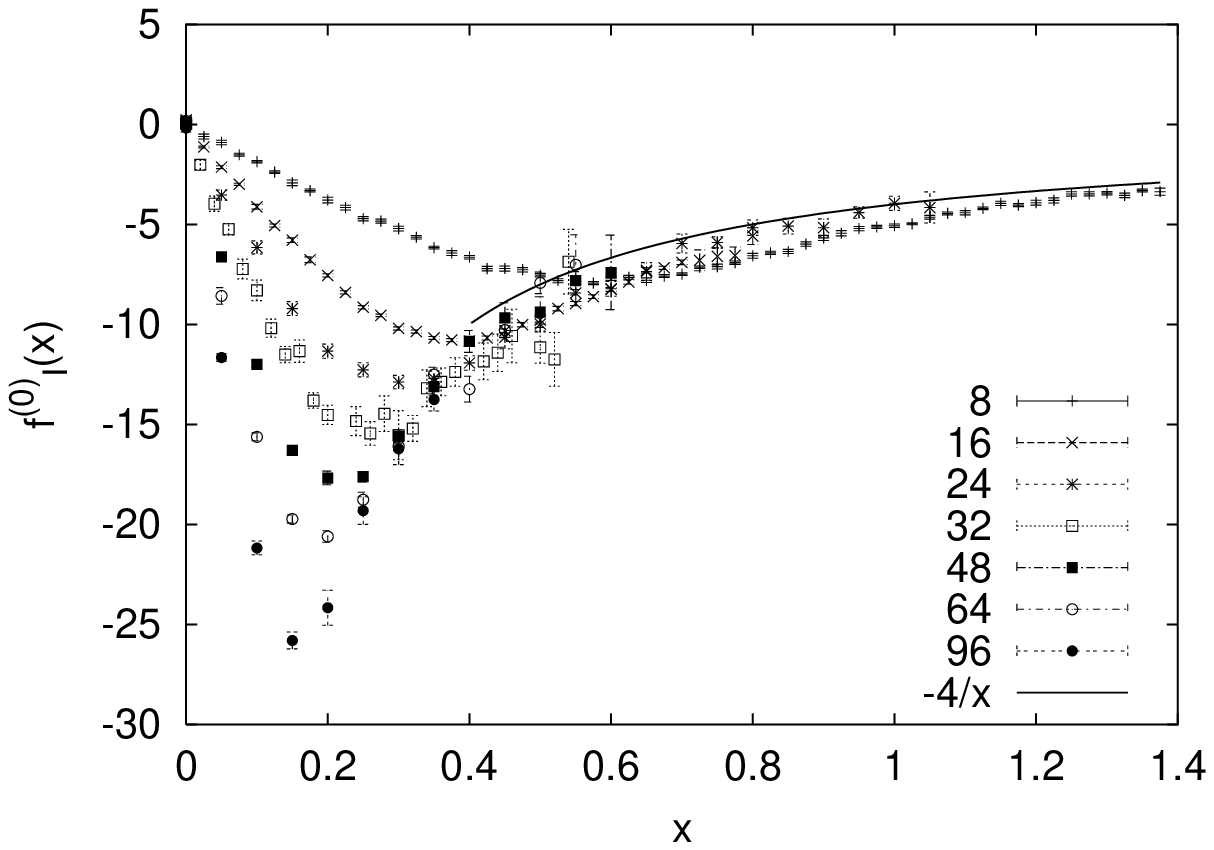}
    \caption{The function $f _{\rm I} ^{(0)} (x)$ is plotted
% against $x$
for $N=8,16,24,32,48,64,96$ at $\mu = 0.2$.
The solid line represents the asymptotic behavior (\ref{f0I_fit})
discussed in the Appendix.
}
    \label{fig:f0I}
  \end{center}
\end{figure}

\begin{figure}[htbp]
  \begin{center}
    \includegraphics[height=8cm]{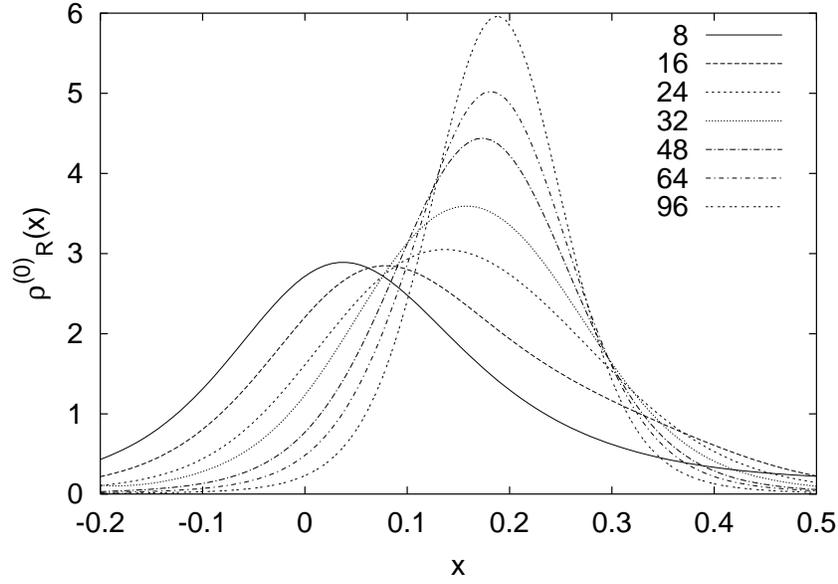}
    \caption{The function $\rho _{\rm R} ^{(0)} (x)$
is plotted
% against $x$ 
for $N=8,16,24,32,48,64,96$ at $\mu = 0.2$.
The position of the peak is approaching
the large $N$ result $\vev{\nu_{\rm R}}_0 = 0.2$ for the
phase quenched system.
}
    \label{fig:rho0R}
  \end{center}
\end{figure}

\begin{figure}[htbp]
  \begin{center}
    \includegraphics[height=8cm]{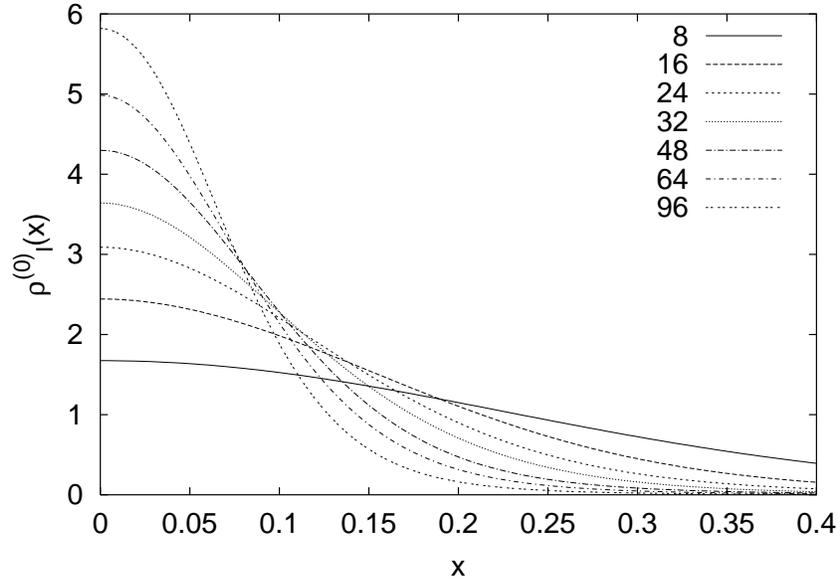}
    \caption{The function $\rho _{\rm I} ^{(0)} (x)$
is plotted 
%against $x$
for $N=8,16,24,32,48,64,96$ at $\mu = 0.2$.
}
    \label{fig:rho0I}
  \end{center}
\end{figure}

\begin{figure}[htbp]
  \begin{center}
    \includegraphics[height=8cm]{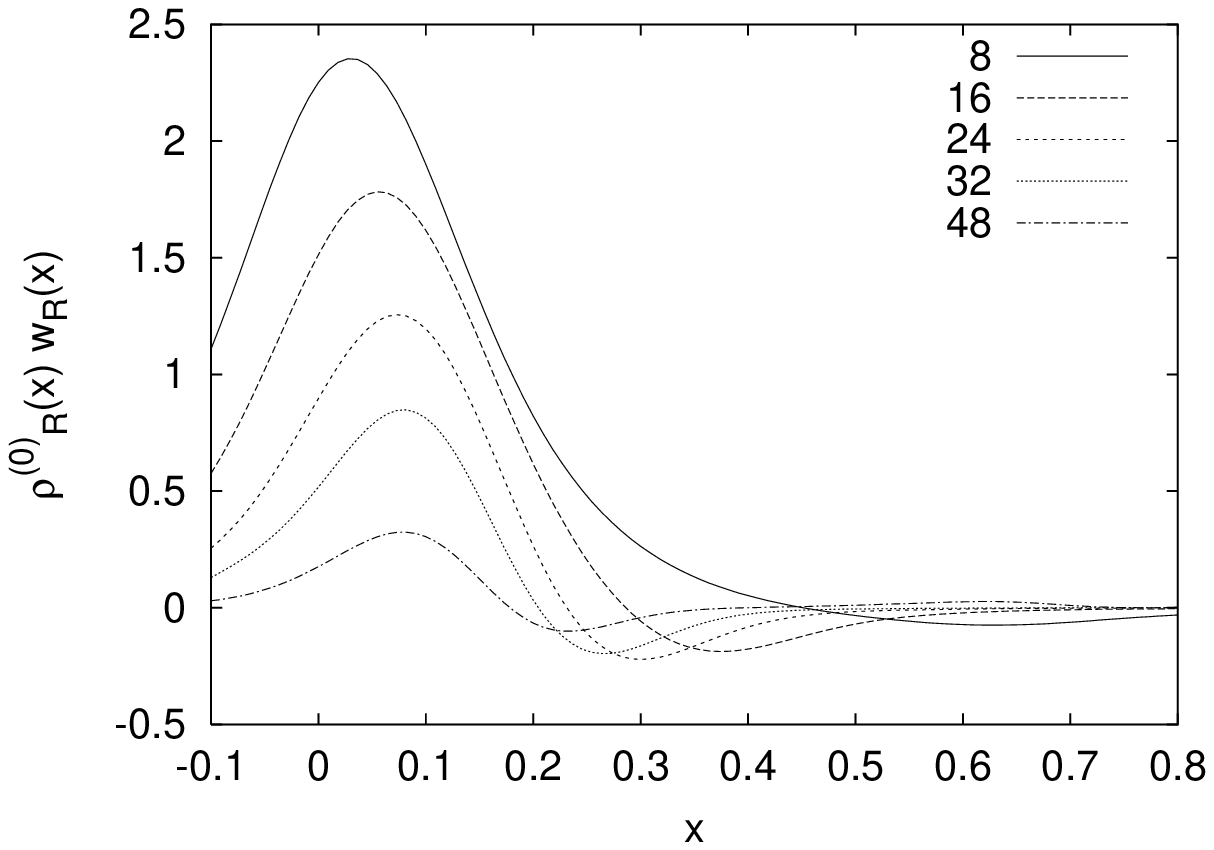}
    \caption{The product $\rho _{\rm R} ^{(0)} (x)w_{\rm R}(x)$ 
is plotted against $x$ for $N=8,16,24,32,48$ at $\mu = 0.2$.}
    \label{fig:rhowR}
  \end{center}
\end{figure}

\begin{figure}[htbp]
  \begin{center}
    \includegraphics[height=8cm]{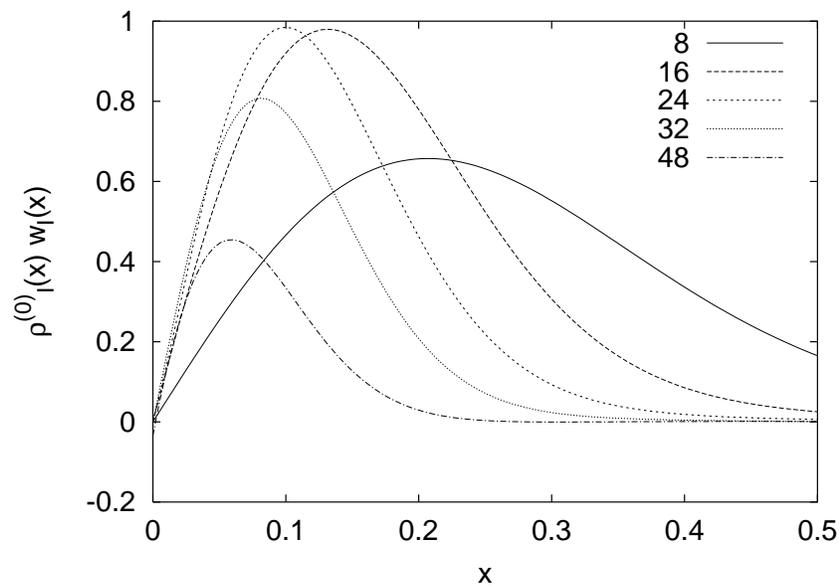}
    \caption{The product $\rho _{\rm I} ^{(0)} (x) w_{\rm I}(x)$
is plotted against $x$
for $N=8,16,24,32,48$ at $\mu = 0.2$.}
    \label{fig:rhowI}
  \end{center}
\end{figure}

We start with the results for $\mu = 0.2$.
In Figs.\ \ref{fig:wR} and \ref{fig:wI}, 
we plot $w_{\rm R}(x)$ and $w_{\rm I}(x)$ respectively.
%The absolute value $|w_i(x)|$
%becomes smaller as $N$ increases, but we were able to 
%obtain results up to $N=64$ with reasonable efforts.
%
%We also note that 
%$w_{\rm R}(x)$ changes its sign at some $x$.
In Figs.\ \ref{fig:f0R} and \ref{fig:f0I},
we show the results for the functions $f_i^{(0)} (x)$.
(The behavior of these functions can be understood theoretically
as discussed in the Appendix.)
By integrating these functions and exponentiating,
we obtain the $\rho_i^{(0)} (x)$, which is shown in 
Figs.\ \ref{fig:rho0R} and \ref{fig:rho0I}.
In Fig.\ \ref{fig:rhowR} and \ref{fig:rhowI},
we show $\rho _{\rm R} ^{(0)} (x)w_{\rm R}(x)$ 
and $\rho _{\rm I} ^{(0)} (x) w_{\rm I}(x)$,
respectively.
Using them we obtain $\langle \nu_{\rm R} \rangle $ 
and $i \, \langle \nu_{\rm I}\rangle $.
Summing these values, we get
$\langle \nu \rangle$,
which should be compared with the exact result obtained from Eq.\
(\ref{finiteN}). The results are shown in Table\ \ref{t:1}.

Note that the sign change of $w_{\rm R}(x)$
occurs near the peak of $\rho _{\rm R} ^{(0)} (x)$,
so that the product 
$\rho _{\rm R} ^{(0)} (x)w_{\rm R}(x)$ 
has a positive regime and a negative regime, which cancel
each other resulting in  $\langle \nu_{\rm R} \rangle \sim 0$.
Thus the main contribution to $\langle \nu \rangle$ comes from the 
imaginary part $\langle \nu_{\rm I} \rangle$ in contrast to 
the results (\ref{RI0}) for the phase quenched system.
This is consistent with the theoretical argument 
at the end of Section \ref{problem}
that the contribution to $\langle \nu \rangle$ comes {\em solely}
from the imaginary part in the large $N$ limit.
%As already mentioned in section 3, this result can be interpreted
%as follows: if we calculate the baryon density at $-\mu$ using
%a partition function for chemical potential $+\mu$, the phased quenched
%approximation agrees with the unquenched result in the thermodynamic limit.
%The difference between the phase quenched and unquenched results
%is mainly through the cancellation
%of poles of the baryon density operator by the fermion determinant. % jac

%\newpage

\subsection{$\mu > \mu_{\rm c}$}

\begin{figure}[htbp]
  \begin{center}
    \includegraphics[height=8cm]{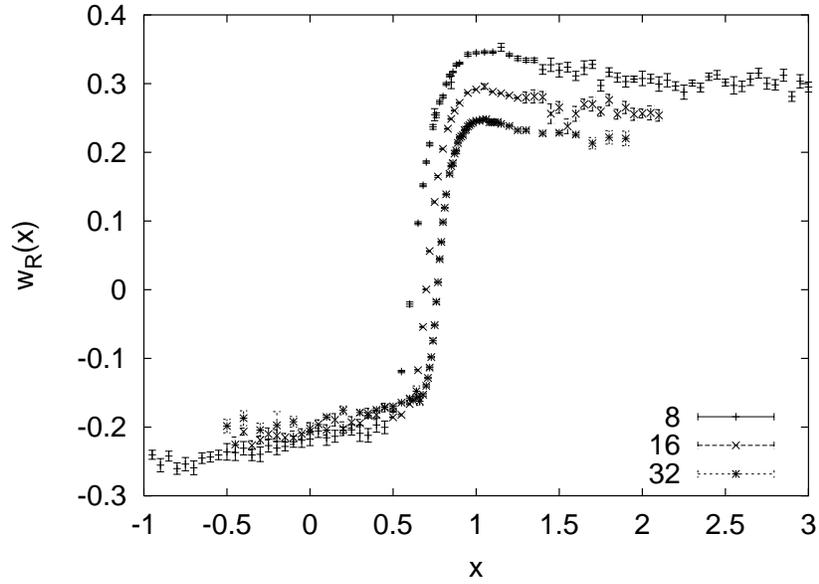}
    \caption{The weight factor $w_{\rm R}(x)$ is plotted against $x$
for $N=8,16,32$ at $\mu = 1.0$.}
    \label{fig:wR_mu1}
  \end{center}
\end{figure}

\begin{figure}[htbp]
  \begin{center}
    \includegraphics[height=8cm]{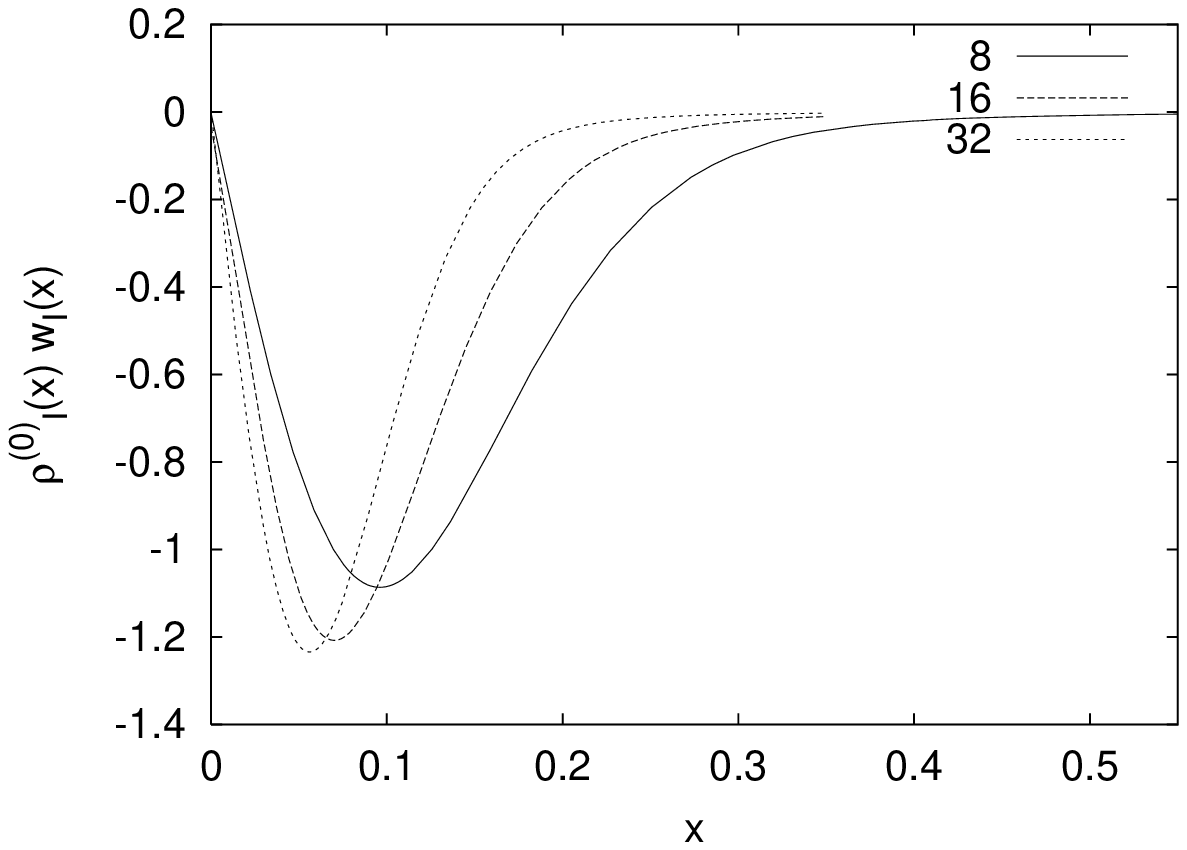}
    \caption{The product $\rho _{\rm I} ^{(0)} (x)w_{\rm I}(x)$ 
is plotted against $x$ for $N=8,16,32$ at $\mu = 1.0$.}
    \label{fig:rhowI_mu1}
  \end{center}
\end{figure}

We perform a similar analysis at $\mu = 1.0$.
The results are presented in 
Figs.\ \ref{fig:wR_mu1}, \ldots , \ref{fig:rhowI_mu1}.
Again the results nicely reproduce the exact results
as can be seen in Table\ \ref{t:1}.
Note in particular that the finite $N$ effect is 1.6\%
at $N=32$, meaning that the accessible values of $N$ are 
large enough to extract the large $N$ limit.

Here, $w_{\rm R}(x)$ is approximately constant
in the region where $\rho _{\rm R} ^{(0)} (x)$ is peaked,
so the shape of the product 
$\rho _{\rm R} ^{(0)} (x)w_{\rm R}(x)$ 
is similar to $\rho _{\rm R} ^{(0)} (x)$.
On the other hand, the peak of $\rho _{\rm I} ^{(0)} (x)$
at $x=0$ is slightly shifted by multiplying $w_{\rm I}(x)$, 
%which becomes zero at $x=0$,
but the first moment of the product
$\rho _{\rm I} ^{(0)} (x)w_{\rm I}(x)$ is still small.
This is also the case for $\mu = 0.2$, but
the difference comes from the normalization constant $C$ in the formula
(\ref{simpleformula}) for $\langle \nu_{\rm I} \rangle$.
As in (\ref{C_new}), the constant $C$ is obtained by 
integrating $\rho _{\rm R} ^{(0)} (x)w_{\rm R}(x)$,
where cancellations occur at $\mu = 0.2$,
but {\em not} at $\mu = 1.0$.
As a result we obtain 
$\langle \nu_{\rm I} \rangle \sim 0$ at $\mu = 1.0$ as $N$ increases.
Thus the main contribution to $\langle \nu \rangle$ comes from the 
real part $\langle \nu_{\rm R} \rangle$,
and moreover, it is close to $\langle \nu_{\rm R} \rangle_0$.
Again this is consistent with the theoretical argument
given at the end of Section \ref{problem}.
%The interpretation is that the phase quenched approximation gives
%the correct result in the thermodynamic limit 
%for $\mu $ outside the domain of 
%eigenvalues of $W$ which is the complex unit circle.
%
% The interpretation is that the phase quenched approximation gives
% the correct result in the thermodynamic limit and $\mu > \mu_{\rm c}$.
% Indeed, this was expected for $\mu $ outside the domain of 
% eigenvalues of $W$ which is the complex unit circle, but that
% this already holds beyond $\mu = \mu_{\rm c}$ is less intuitive and has
% only been understood in terms of the exact result for the Random
% Matrix Model.
% On the other hand, $\rho _{\rm I} ^{(0)} (x)$
% is peaked around $x=0$, which is shifted in
% $\rho _{\rm I} (x)$ due to the multiplication
% by $w_{\rm I}(x)$.
% becomes zero due to symmetry.
% As a result $\rho _{\rm I} ^{(0)} (x)w_{\rm I}(x)$ 
% is much smaller than 
% $\rho _{\rm R} ^{(0)} (x)w_{\rm R}(x)$ 
% in magnitude, and we obtain 
% $\langle \nu_{\rm I} \rangle \sim 0$.
%It is clear from the Table that the $N$ 

It is interesting that
the $w_{\rm R}(x)$ changes from positive to negative for $\mu =0.2$,
but it changes from negative to positive for $\mu =1.0$.
Similarly $w_{\rm I}(x)$ is positive at $x>0$ for $\mu =0.2$,
but it is negative at $x>0$ for $\mu =1.0$.
Thus the behavior of $w_i (x)$ changes drastically as 
the chemical potential $\mu$ crosses its critical value $\mu_{\rm c}$.
%are quite different depending on the phase.
%Note in this regard that the function $\rho ^{(0)}_i(x)$
%changes smoothly as $\mu$ crosses $\mu_{\rm c}$.

\subsection{$\mu \sim \mu_{\rm c}$}

\begin{figure}[htbp]
  \begin{center}
    \includegraphics[height=8cm]{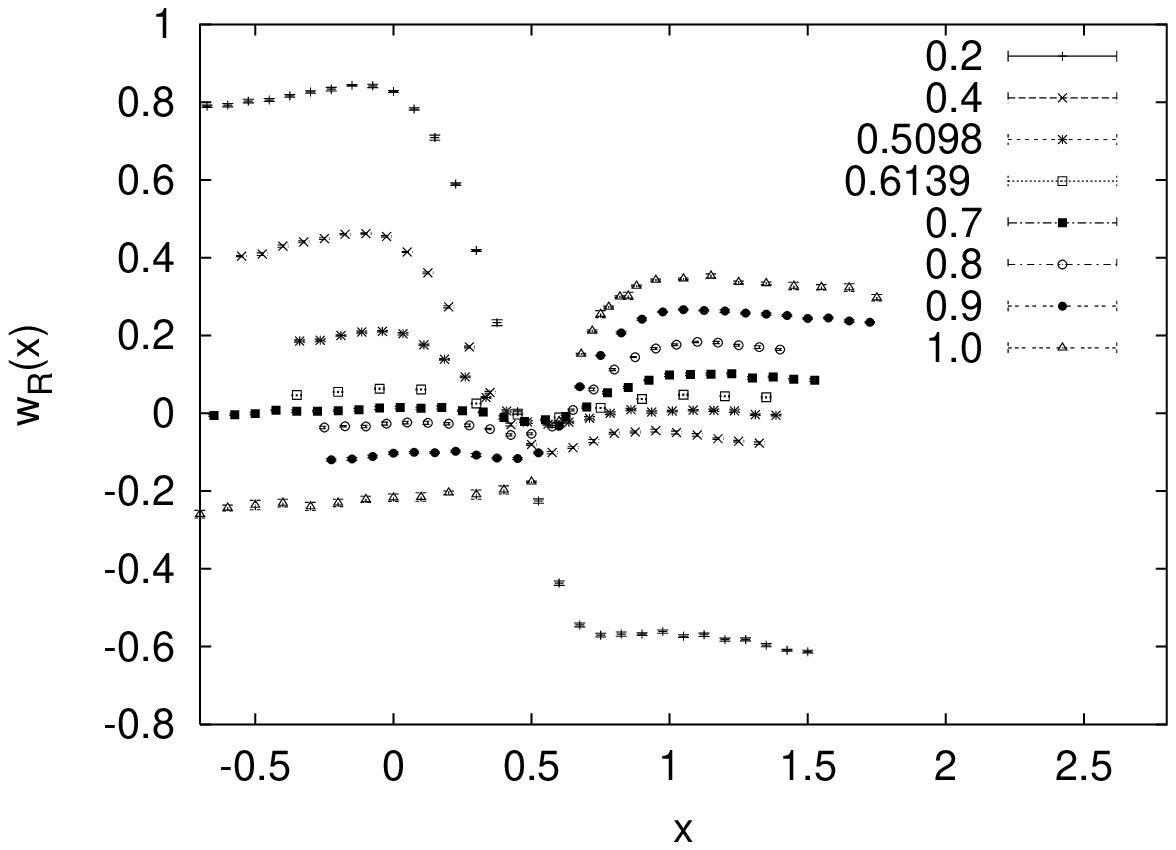}
    \caption{The weight factor $w_{\rm R}(x)$ is plotted 
against $x$ 
for $N=8$ at various $\mu$.
The behavior changes drastically as $\mu$ crosses 
the critical point.}
    \label{fig:wR_N8}
  \end{center}
\end{figure}

\begin{figure}[htbp]
  \begin{center}
    \includegraphics[height=8cm]{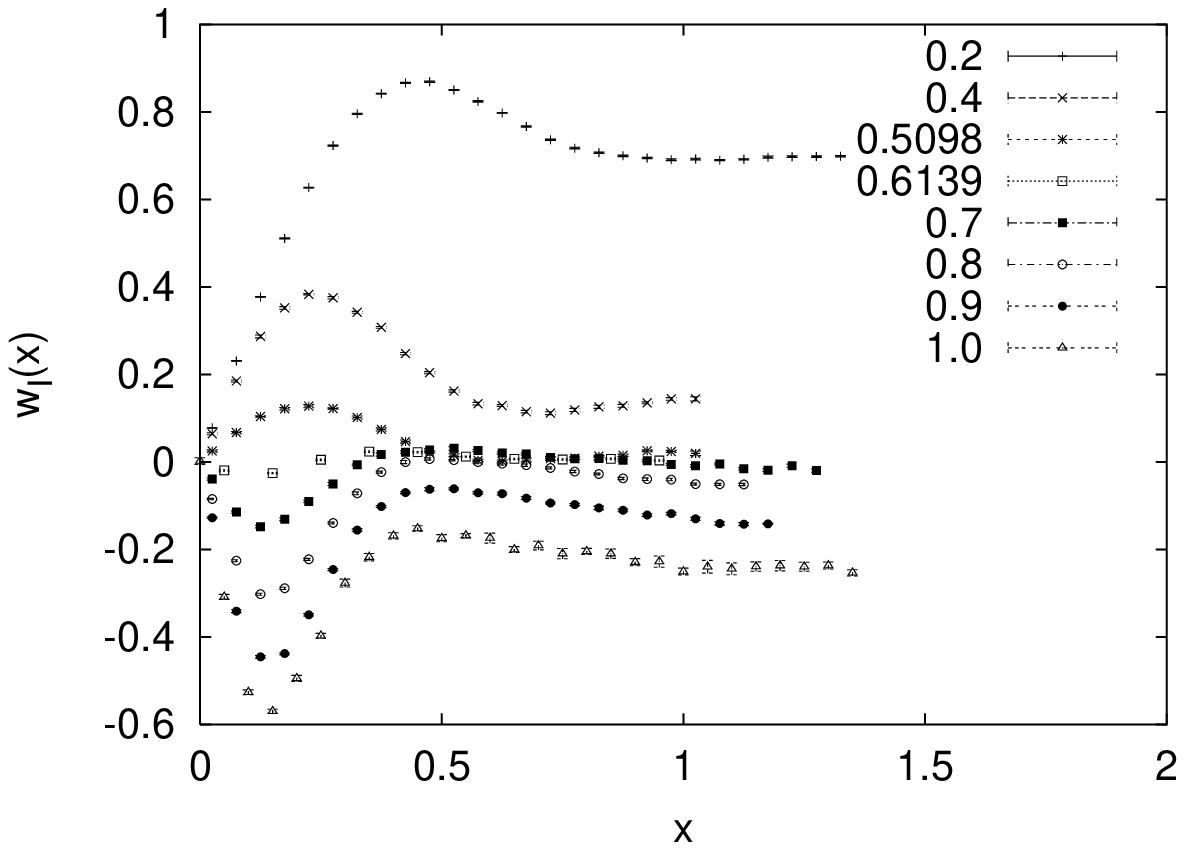}
    \caption{The weight factor $w_{\rm I}(x)$ is plotted 
against $x$ for $N=8$ at various $\mu$.
The behavior changes drastically as $\mu$ crosses 
the critical point.
}
    \label{fig:wI_N8}
  \end{center}
\end{figure}

\begin{figure}[htbp]
  \begin{center}
    \includegraphics[height=8cm]{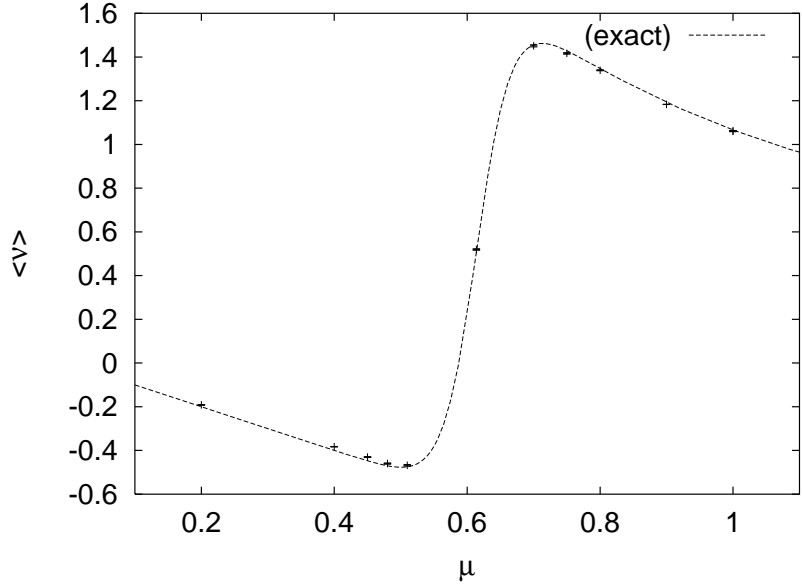}
    \caption{The VEV $\vev{\nu}$ is obtained by the factorization
method at $N=8$ for various $\mu$ including the critical regime. 
%including the critical regime. 
Statistical errors computed by the jackknife method are also shown.
The dashed line represents the exact result 
(\protect\ref{finiteN}) for $\vev{\nu}$ at $N=8$.}
    \label{fig:crit_N8}
  \end{center}
\end{figure}

Let us see in more detail what is going on
in the critical regime.
In Figs.\ \ref{fig:wR_N8} and \ref{fig:wI_N8}
we plot $w_{\rm R}(x)$ and $w_{\rm I}(x)$ respectively for $N=8$
at various $\mu$.
The final results for $\vev{\nu_i}$ are plotted
in Fig.\ \ref{fig:crit_N8}
and the corresponding data are listed in
Table \ref{t:2}.
Note that we were able to reproduce the exact result
even at the turning point ($\mu = 0.6139$).
% for $N=8$.
%which is considered to be the most difficult point.
%Thus for $N=8$ we 
%
%In the critical regime, the weight functions
%$w_i(x)$ becomes very small in magnitude,
%and therefore (\ref{rel0})$\sim$(\ref{rel2}) becomes small.
%(\ref{reweight})
These results provide a clear understanding of 
how the first order phase transition occurs
due to the effects of $\Gamma$.
%, namely the imaginary part of the action.

\begin{figure}[htbp]
  \begin{center}
    \includegraphics[height=8cm]{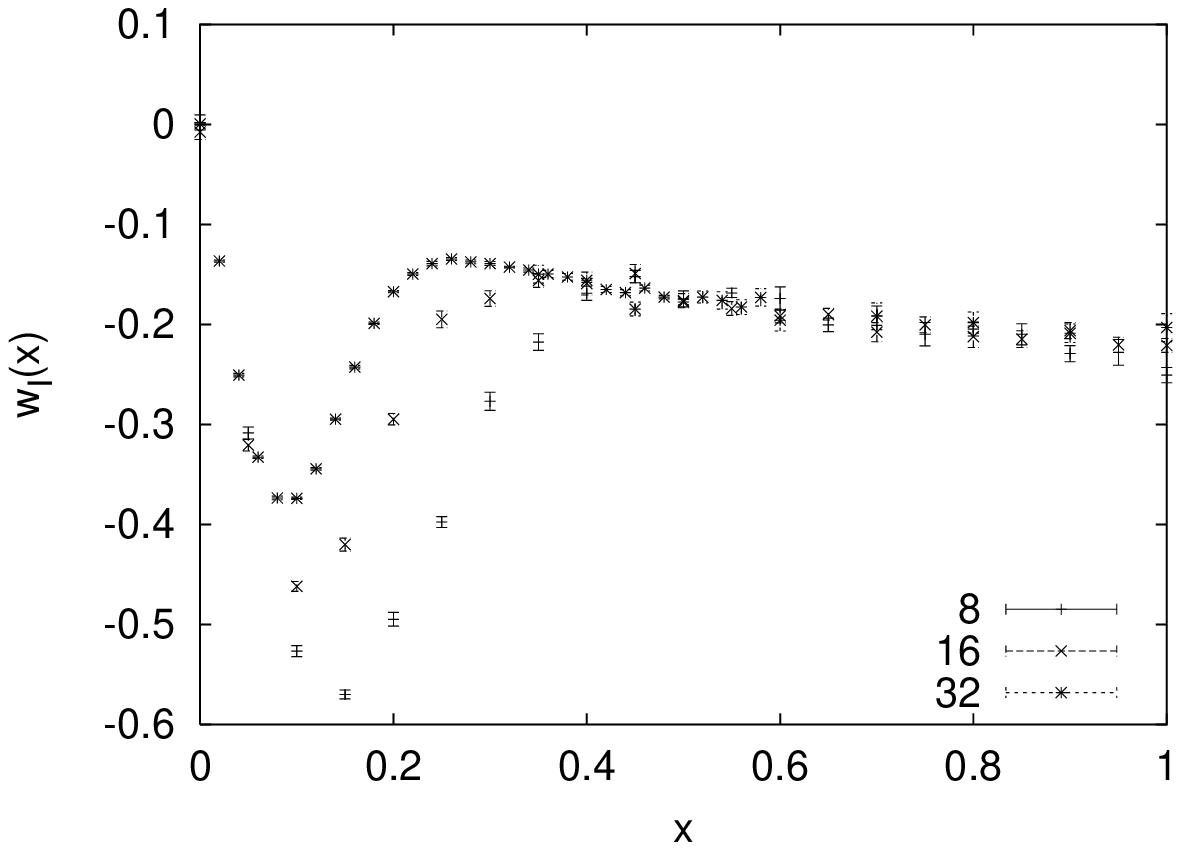}
    \caption{The weight factor $w_{\rm I}(x)$ 
is plotted against $x$
for $N=8,16,32$ at $\mu = 1.0$.}
    \label{fig:wI_mu1}
  \end{center}
\end{figure}

\begin{figure}[htbp]
  \begin{center}
    \includegraphics[height=8cm]{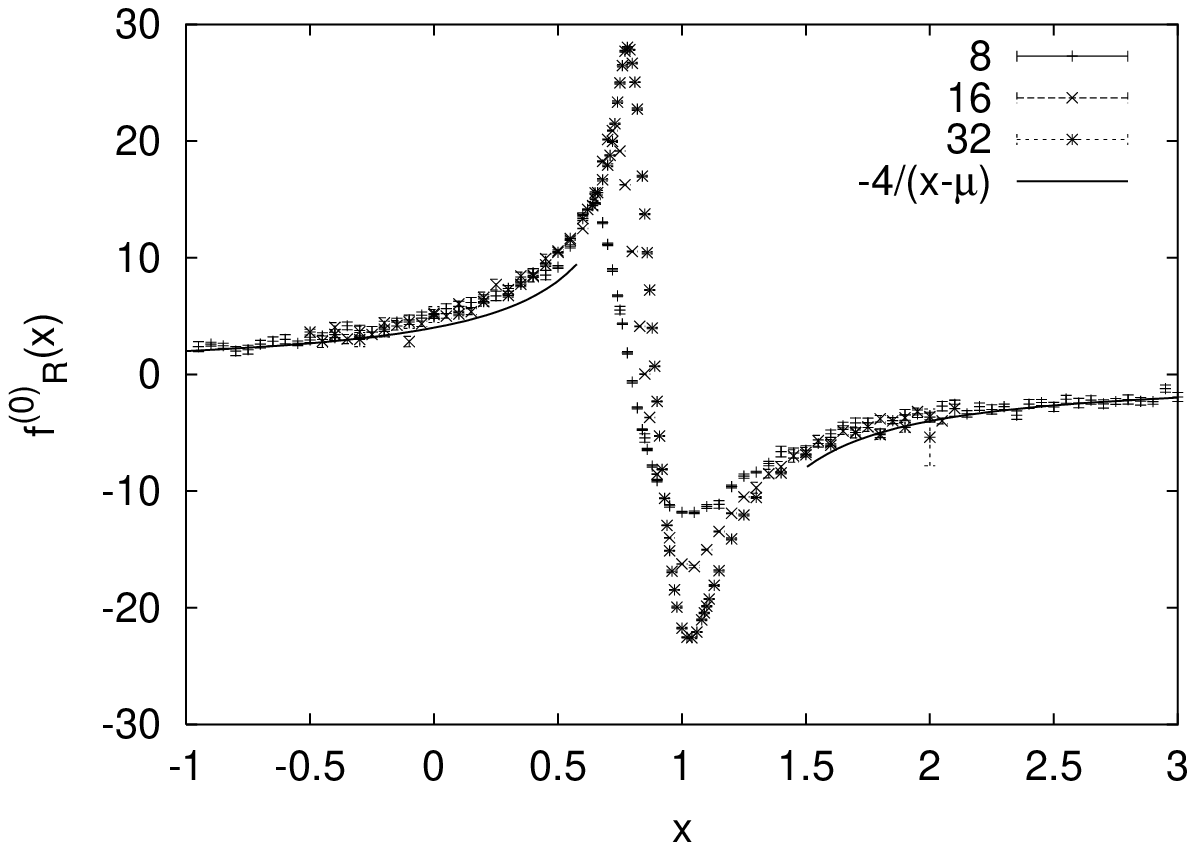}
    \caption{The function $f _{\rm R} ^{(0)} (x)$ is plotted
% against $x$
for $N=8,16,32$ at $\mu = 1.0$.
The solid line represents the asymptotic behavior (\ref{f0R_fit})
discussed in the Appendix.
}
    \label{fig:f0R_mu1}
  \end{center}
\end{figure}

\begin{figure}[htbp]
  \begin{center}
    \includegraphics[height=8cm]{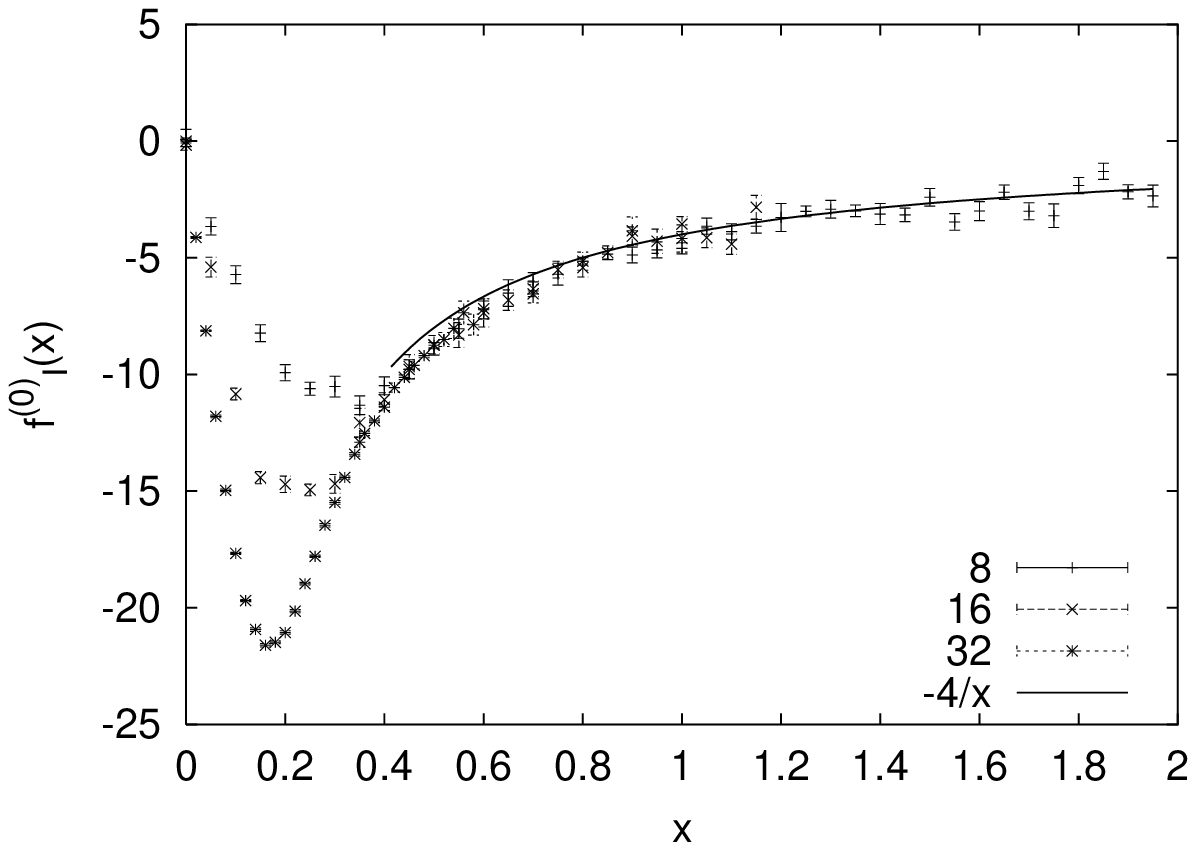}
    \caption{The function $f _{\rm I} ^{(0)} (x)$ is plotted
% against $x$
for $N=8,16,32$ at $\mu = 1.0$.
The solid line represents the asymptotic behavior (\ref{f0I_fit})
discussed in the Appendix.
}
    \label{fig:f0I_mu1}
  \end{center}
\end{figure}

\begin{figure}[htbp]
  \begin{center}
    \includegraphics[height=8cm]{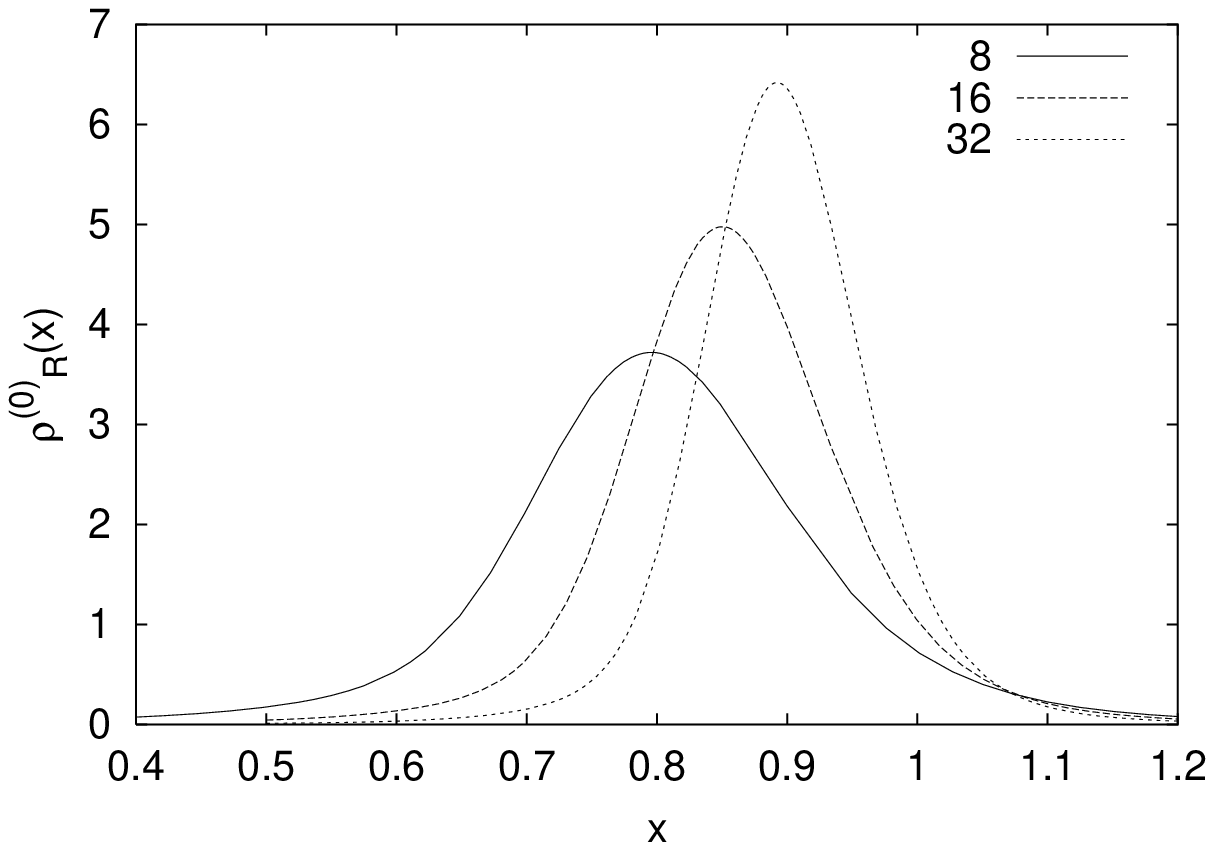}
    \caption{The function $\rho _{\rm R} ^{(0)} (x)$
is plotted 
%against $x$ 
for $N=8,16,32$ at $\mu = 1.0$.
The position of the peak is approaching
the large $N$ result $\vev{\nu_{\rm R}}_0 = 1$ for the
phase quenched system.
}
    \label{fig:rho0R_mu1}
  \end{center}
\end{figure}

\begin{figure}[htbp]
  \begin{center}
    \includegraphics[height=8cm]{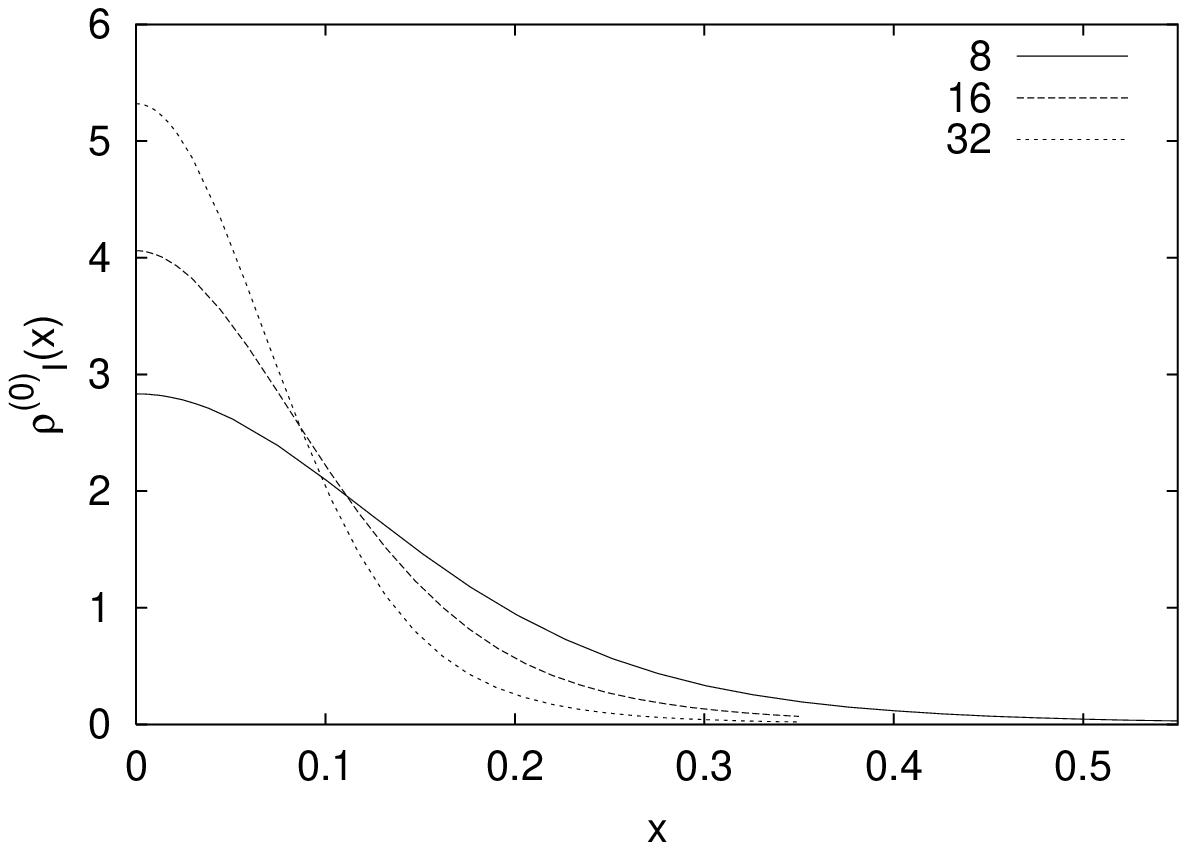}
    \caption{The function $\rho _{\rm I} ^{(0)} (x)$
is plotted 
%against $x$
for $N=8,16,32$ at $\mu = 1.0$.}
    \label{fig:rho0I_mu1}
  \end{center}
\end{figure}

\begin{figure}[htbp]
  \begin{center}
    \includegraphics[height=8cm]{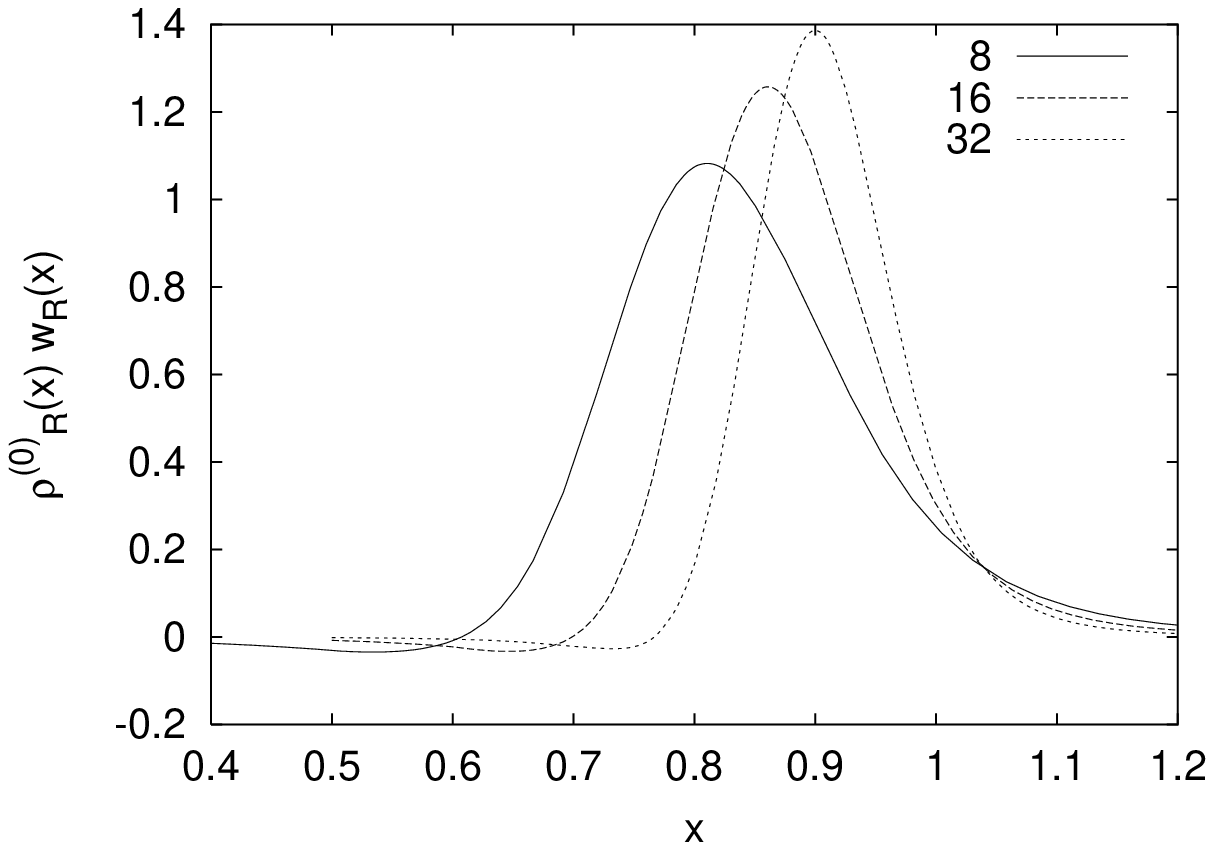}
    \caption{The product $\rho _{\rm R} ^{(0)} (x)w_{\rm R}(x)$ 
is plotted against $x$ for $N=8,16,32$ at $\mu = 1.0$.}
    \label{fig:rhowR_mu1}
  \end{center}
\end{figure}

\begin{table}[ht]
\begin{center}
\begin{tabular}{|c | c | c | c | c | c | }
\hline\hline
$\mu$&$N$ &$\vev{\nu_{\rm R}}$&$i \, \vev{\nu_{\rm I}}
$&$\vev{\nu}$&$\vev{\nu}$~(exact)\\
\hline
 0.2 &  8 & 0.0056(6)& -0.1970(5)& -0.1915(8)&-0.200000 \\
 0.4 &  8 & 0.0120(7)& -0.395(1) & -0.383(2) &-0.399743 \\
 0.45&  8 & 0.011(1) & -0.442(2) & -0.430(3) &-0.447327 \\
 0.48&  8 & 0.016(2) & -0.476(3) & -0.460(4) &-0.470018 \\
 0.5098&8 & 0.032(2) & -0.500(3) & -0.468(5) &-0.474945 \\
 0.6139&8 & 0.522(2) & -0.002(3) &  0.519(2) & 0.511115 \\
 0.7 &  8 & 0.971(3) &  0.480(4) &  1.451(6) & 1.45516  \\
 0.75&  8 & 0.972(1) &  0.444(3) &  1.417(4) & 1.42899  \\
 0.8 &  8 & 0.949(1) &  0.390(2) &  1.339(3) & 1.34792  \\
 0.9 &  8 & 0.9054(3)&  0.2779(4)&  1.1833(6)& 1.19253  \\
 1.0 &  8 & 0.8622(1)&  0.1999(2)&  1.0620(2)& 1.06650  \\
\hline\hline
\end{tabular}
\end{center}
\caption{Results of the analysis of $\vev{\nu}$ for $N=8$ at various $\mu$
including the critical regime. 
Statistical errors computed by the jackknife method are also shown.
The last column represent the exact result 
(\protect\ref{finiteN}) for $\vev{\nu}$ at $N=8$.
}
\label{t:2}
\end{table}

\begin{figure}[htbp]
  \begin{center}
    \includegraphics[height=8cm]{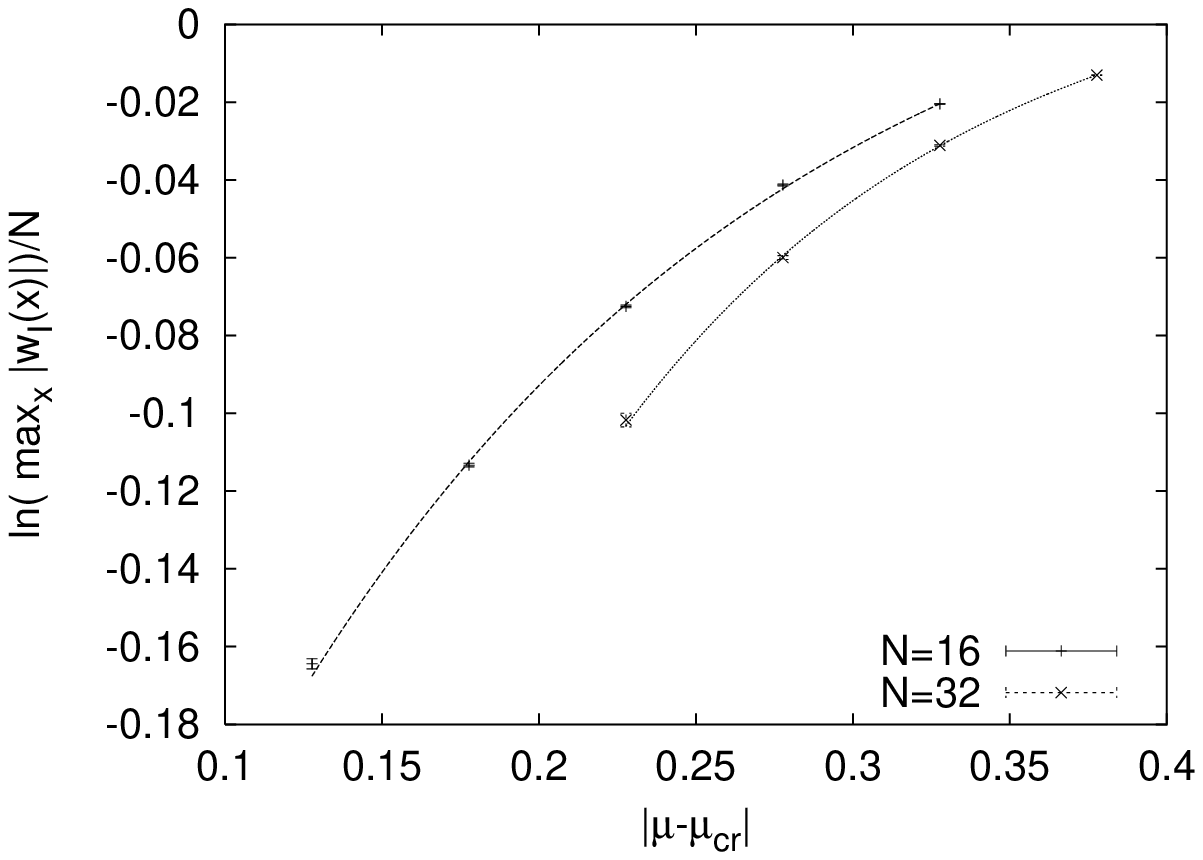}
    \caption{The result of $\ln (\max_x |w_{\rm I}(x)|)$
is plotted against $|\mu-\mu_{\rm c}|$ for $N=16,32$ 
at $\mu < \mu_{\rm c}$.
The lines are the fits to the behavior (\ref{critical_reg}).}
    \label{fig:critical_smallmu}
  \end{center}
\end{figure}

\begin{figure}[htbp]
  \begin{center}
    \includegraphics[height=8cm]{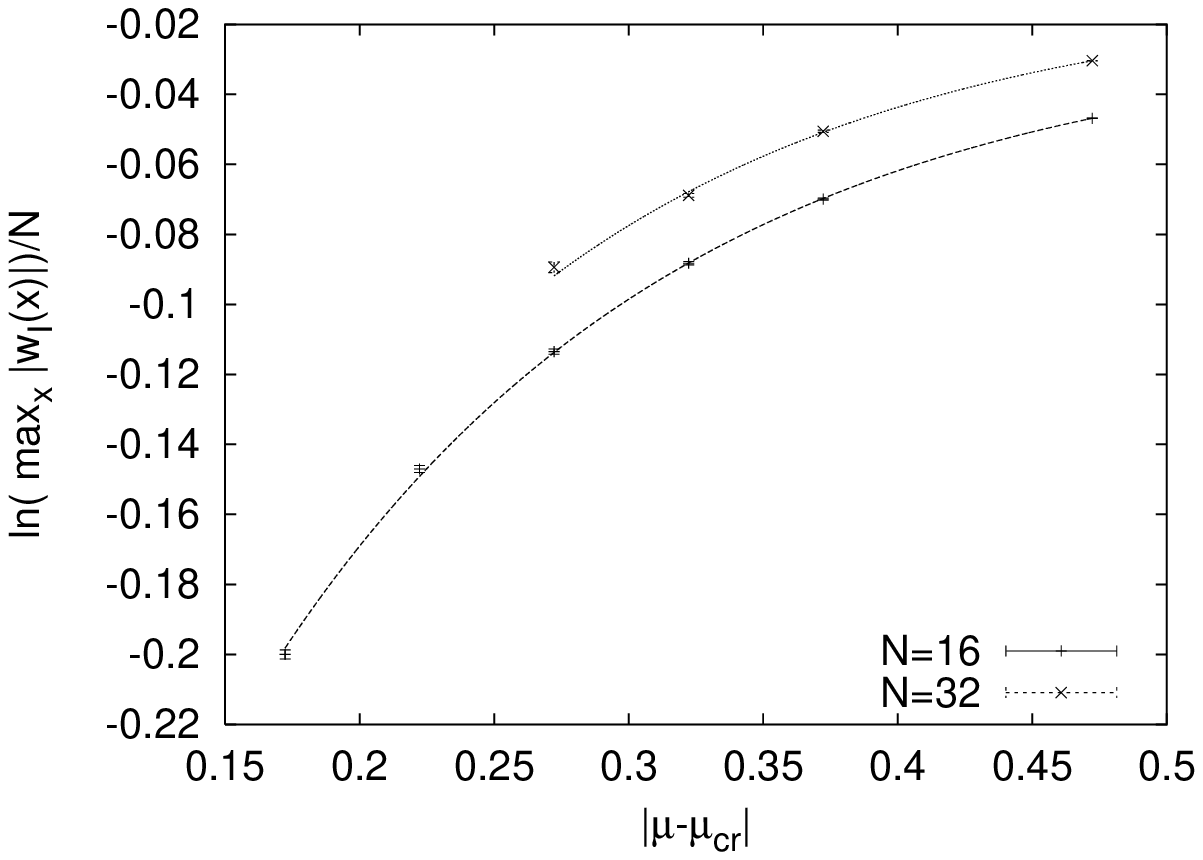}
    \caption{The result of $\ln (\max_x |w_{\rm I}(x)|)$
is plotted against $|\mu-\mu_{\rm c}|$ for $N=16,32$ 
at $\mu > \mu_{\rm c}$.
The lines are the fits to the behavior (\ref{critical_reg}).
}
    \label{fig:critical_largemu}
  \end{center}
\end{figure}

\begin{figure}[htbp]
  \begin{center}
    \includegraphics[height=8cm]{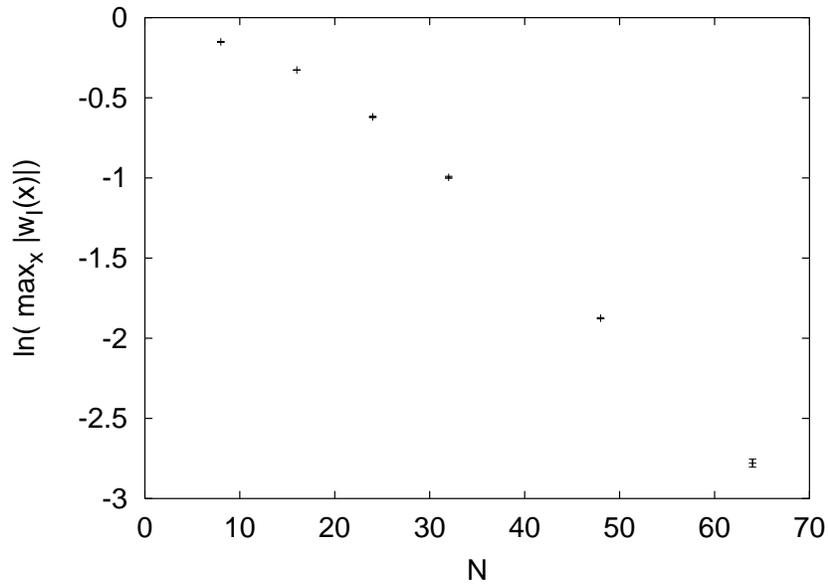}
    \caption{The result of $\ln (\max_x |w_{\rm I}(x)|)$ 
at $\mu = 0.2$
is plotted against $N$.
A linearly decreasing behavior is observed at large $N$.
}
    \label{fig:exp_decay}
  \end{center}
\end{figure}

%On the other hand,
The fact that $|w_i(x)|$ becomes small
near the critical regime reveals an increasing difficulty
in approaching the critical point.
In Figs.\ \ref{fig:critical_smallmu},
\ref{fig:critical_largemu},
we plot $\ln (\max_x |w_{\rm I}(x)|)$ 
against $|\mu - \mu_{\rm c}|$ at $N=16,32$ 
for $\mu < \mu_{\rm c}$ and
$\mu > \mu_{\rm c}$ respectively.
Our data can be nicely fitted to
%We find that
\beq
\ln (\max_x |w_{\rm I}(x)|)
=  - a \exp (- b |\mu-\mu_{\rm c}|) + {\rm const}. \ .
\label{critical_reg}
\eeq
%which means that $\max_x |w_{\rm I}(x)|$ becomes rapidly small 
%as one approaches the critical point.
%
%The fitting parameter $\beta$ is given for $N=16$
%as 5.5(7) for $\mu < \mu_{\rm c}$ and
%as 7.3(8) for $\mu > \mu_{\rm c}$.
%For $N=32$, 8.8(5) for $m<m_c$ 7.0(5) for $m>m_c$
%There is not enough data to truly report errors (only statistical
%errors shown), e.g. one cannot paly with the range to test stability of
%the fit. It could be that adding more points close to m_c (if it were
%possible) that the values could change considerably.
%You can report them as such or say that beta is roughly 9 and
%roughly 5 respectively using the available data.
%
For fixed $\mu$, on the other hand, 
we observe that $\max_x |w_{\rm I}(x)|$
decreases exponentially as $\propto\exp(- c N)$.
We plot the result for $\mu = 0.2$ in Fig.\ \ref{fig:exp_decay}.
This is consistent with the 
scaling behavior of the weight function, 
which plays a crucial role in 
the extrapolation \cite{sign}
as we mentioned at the end of Section \ref{virtue}.

%%%%%%%%%%%%%%%%%

\vspace*{1cm}

\section{Summary}
\label{Summary}
\setcounter{equation}{0}

In this article
we have clarified the properties of the factorization method
for systems with a complex action.
This method circumvents the overlap problem, 
and we hope that this feature alone
will make various interesting questions accessible by the present day
computer resources.
Indeed, we were able to reproduce the exact results for the quark
number density in a schematic model for QCD at finite baryon density.
The achieved system size was already large enough to
obtain the thermodynamic limit.
We therefore expect that the factorization method is useful to 
explore the phase diagram of the `real' finite density QCD.
The method itself is quite general, and it can be applied to any
system with a complex action, although the actual gains may depend on 
the system.

We also emphasize that in the case where the distribution functions
turn out to be positive definite, the method can be even more
powerful by utilizing the scaling property of the weight factor.
This extrapolation appeared particularly valuable in the study of
%to satisfy this condition 
spontaneous symmetry breaking in the type IIB matrix model.
Although this is a very interesting problem which would tell
us a great deal about nonperturbative string dynamics
and the dynamical origin of the space-time dimensionality,
we expect that this method will be useful for other systems as well.

%\vspace*{5mm}

\section*{Acknowledgments}
We would like to thank
W.\ Bietenholz, Z.\ Burda, S.\ Iso, Y.\ Kikukawa, E.\ Kiritsis, 
T.\ Onogi and B.\ Peterssen for discussions.
The authors are also grateful to I.\ Kostov, P.\ Di Francesco
and other organizers of the Sixth Claude Itzykson Meeting
(Matrix Models 2001), which triggered the current 
collaboration.  
K.N.A.'s research was partially supported by RTN grants
HPRN-CT-2000-00122, HPRN-CT-2000-00131 and HPRN-CT-1999-00161 and the
INTAS contract N 99 0590.
The work of J.N.\ was supported in part by Grant-in-Aid for 
Scientific Research (No.\ 14740163) from 
the Ministry of Education, Culture, Sports, Science and Technology.
The work of J.J.M.V.\ was
partially supported by the US DOE grant DE-FG-88ER40388.

\vspace*{1cm}

\section*{Appendix: Large $N$ behavior of $\rho^{(0)}_i (x)$}
\begin{figure}[htbp]
  \begin{center}
    \includegraphics[height=8cm]{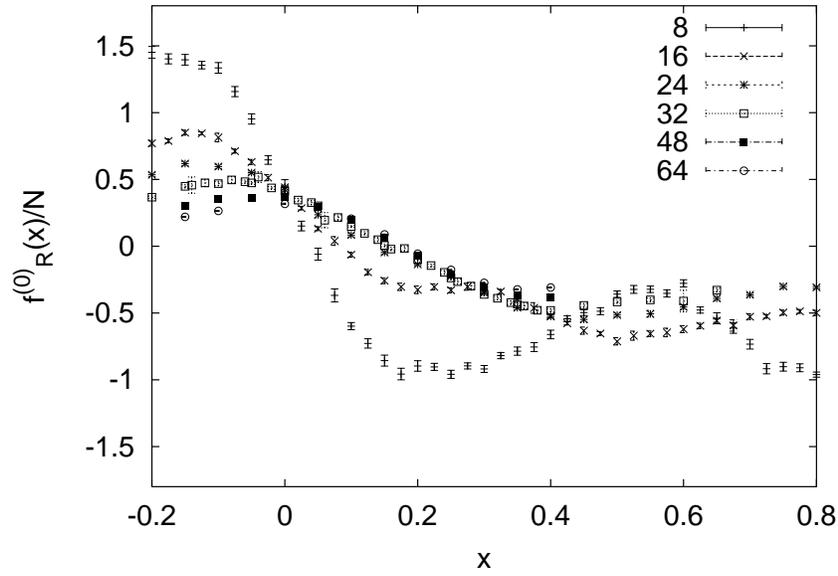}
    \caption{The function $\frac{1}{N}f^{(0)}_{\rm R}(x)$ is plotted
    for $\mu = 0.2$. A clear scaling behavior is seen in the linear
    regime, where the function crosses zero.}
    \label{fig:f0R_scale}
  \end{center}
\end{figure}

\begin{figure}[htbp]
  \begin{center}
    \includegraphics[height=8cm]{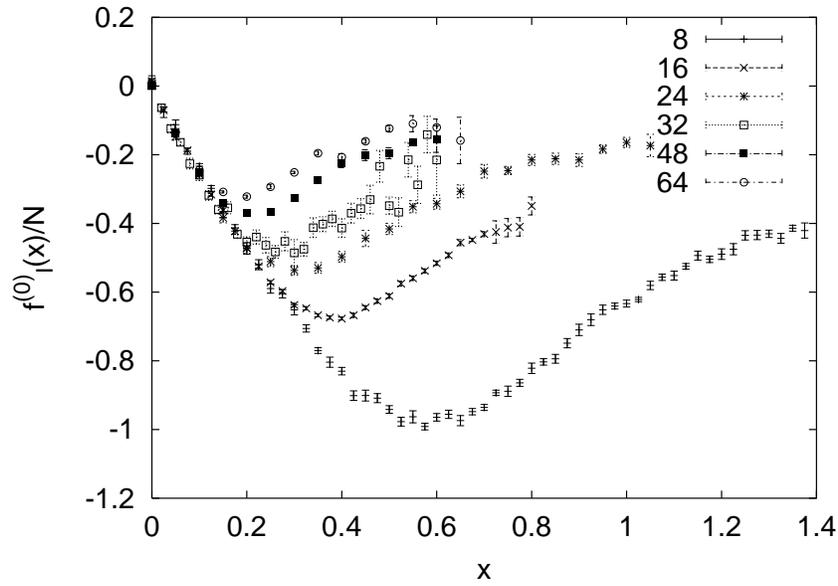}
    \caption{The function $\frac{1}{N}f^{(0)}_{\rm I}(x)$ is plotted
    for $\mu = 0.2$. A clear scaling behavior is seen in the linear
    regime, where the function crosses zero.}
    \label{fig:f0I_scale}
  \end{center}
\end{figure}

%\label{section:rho0}
\setcounter{equation}{0}
\renewcommand{\theequation}{A.\arabic{equation}}
\hspace*{\parindent}
In this Appendix, we discuss the large $N$ behavior
of functions $\rho^{(0)}_i (x)$.
From Figures \ref{fig:f0R}, \ref{fig:f0I}, \ref{fig:f0R_mu1},
\ref{fig:f0I_mu1}, we find that 
$\rho^{(0)}_i (x)$ is well approximated by the Gaussian distribution 
near the peak, but there is a transition to a power-like tail 
($\rho^{(0)}_i(x)\propto x^{-4}$) at large $|x|$. 
The function $f^{(0)}_i(x)$ scales in this power regime.
In the Gaussian regime, 
on the other hand, the function 
$\frac{1}{N}f^{(0)}_i(x)$, with the
normalization factor $1/N$, scales as is shown
in Figs.\ \ref{fig:f0R_scale} and \ref{fig:f0I_scale}
for $\mu = 0.2$.
As a result the extent of the Gaussian regime shrinks as $1/\sqrt{N}$.
%-jac
Such a behavior is obtained if $N$ independently distributed eigenvalues
of $W$ contribute to $\langle \nu \rangle$. 
Indeed, the correlations of the eigenvalues of the matrix $W$ in our
model decay exponentially on the scale
of the average level spacing.
%In Figs.\ \ref{fig:f0R_scale} and \ref{fig:f0I_scale}
%we plot $\frac{1}{N}f^{(0)}_i(x)$ for $\mu = 0.2$ which shows that
%the behavior of the function $\rho^{(0)}_i (x)$ near the peak is given
%by
%\beq
%\rho ^{(0)}_i (x) \propto \sqrt N \ee^{N \Omega_i (x)} \ ,
%\label{rho0scale}
%\eeq
%where $\Omega_i(x)$ is an $N$-independent function of $x$.
%Away from the peak, the functions $\rho^{(0)}_i (x)$ without 
%the factor $1/N$ scales.
%%This means that the effects of the phase becomes dominant at large
%%$N$.

The observed power tail can be understood as follows.
The point is that large values of the baryon number result
from eigenvalues of the matrix $W$ 
close to $\pm i\mu$. The probability of finding one
eigenvalue $\lambda$ 
inside a radius $\alpha$ can be easily obtained for $\mu = 0$
from generalizing a calculation in the book of Mehta \cite{mehtabook}
(the chapter on
RMT's without hermiticity conditions). One finds that 
\beq
P(\exists\lambda < \alpha) = \frac {N^2}2 \alpha^4
\quad {\rm for} \quad \alpha \to 0 \ .
\eeq
If we assume stationarity of this result we obtain
\beq
P(|\nu| > x) = P\left(\exists \lambda , |\lambda-i\mu| < \frac 1{2Nx}\right)
= \frac 1{32N^2 x^4} \ .
\eeq
For the probability density we thus find
\beq
P(|\nu| = x) = \frac 1{8 N^2 x^5} \ ,
\eeq
so that the real and imaginary parts of the quark number density behave
as $1/x^4$ for large $|x|$.
Since the level spacing distribution is a universal feature of
Random Matrix Theories, we expect that such power-like tails are
generic and also occur in QCD.

In fact our results in 
Figures \ref{fig:f0R}, \ref{fig:f0I}, \ref{fig:f0R_mu1},
\ref{fig:f0I_mu1} suggest that
$f^{(0)}_{\rm R} (x)$ and $f^{(0)}_{\rm I} (x)$ are given at large $|x|$
by
\beqa
\label{f0R_fit}
f^{(0)}_{\rm R} (x)&\sim& -\frac{4}{x-\mu} \\
f^{(0)}_{\rm I} (x)&\sim& -\frac{4}{x} \ ,
\label{f0I_fit}
\eeqa
%For the distribution function we thus have a crossover from a Gaussian
%behavior to a power behavior. 
which agrees with the above argument.

\end{document}